\documentclass[journal]{IEEEtran}

\usepackage{graphics, array, url, hyperref, epsfig, amsmath, amssymb, graphicx}
\usepackage{multirow}
\usepackage{subfigure}

\begin{document}

\title{Enabling Strong Privacy Preservation and Accurate Task Allocation for Mobile Crowdsensing}

\author{Jianbing Ni, \emph{Student Member, IEEE,} Kuan Zhang, \emph{Member, IEEE,} Qi Xia, Xiaodong Lin, \emph{Fellow, IEEE,} \\Xuemin (Sherman) Shen, \emph{Fellow, IEEE} \thanks{Part of this research work was presented in IEEE International Conference on Communications (ICC 2017) \cite{NIICC17}.} \thanks{Jianbing Ni and Xuemin (Sherman) Shen are with Department of Electrical and Computer Engineering, University of Waterloo, Waterloo, Ontario, Canada N2L 3G1. email: \{j25ni, sshen\}@uwaterloo.ca.} \thanks{Kuan Zhang is with Department of Electrical and Computer Engineering, University of Nebraska-Lincoln, Omaha, NE 68182 USA. email: kuan.zhang@unl.edu.} \thanks{Xiaodong Lin is with Department of Physics and Computer Science, Wilfrid Laurier University, Waterloo, Ontario, Canada N2L 3C5, email: xlin@wlu.ca.} \thanks{Qi Xia is with Center of Cyber Security, University of Electronic Science and Technology of China, Chengdu, 611731, China, email: xiaqi@uestc.edu.cn.}}

\maketitle

\begin{abstract}
Mobile crowdsensing engages a crowd of individuals to use their mobile devices to cooperatively collect data about social events and phenomena for special interest customers. It can reduce the cost on sensor deployment and improve data quality with human intelligence. To enhance data trustworthiness, it is critical for service provider to recruit mobile users based on their personal features, e.g., mobility pattern and reputation, but it leads to the privacy leakage of mobile users. Therefore, how to resolve the contradiction between user privacy and task allocation is challenging in mobile crowdsensing.
In this paper, we propose SPOON, a strong privacy-preserving mobile crowdsensing scheme supporting accurate task allocation from geographic information and credit points of mobile users. In SPOON, the service provider enables to recruit mobile users based on their locations, and select proper sensing reports according to their trust levels without invading user privacy. By utilizing proxy re-encryption and BBS+ signature, sensing tasks are protected and reports are anonymized to prevent privacy leakage. In addition, a privacy-preserving credit management mechanism is introduced to achieve decentralized trust management and secure credit proof for mobile users. Finally, we show the security properties of SPOON and demonstrate its efficiency on computation and communication.

\vskip 2mm \noindent{\bf Keywords:} Mobile Crowdsensing, Task Allocation, Trust Management, Privacy Preservation.
\end{abstract}

\section{Introduction}
The development of wireless communications and mobile devices triggers the emergence of mobile crowdsensing \cite{Ganti11}, in which user-centric mobile sensing and computing devices, e.g., smartphones, in-vehicle devices and wearable devices, are utilized to sense, collect and process data from the environment. This ``Sensing as a Service" \cite{Hui17} elaborates our knowledge of the physical world by opening up a new door for data collection and sharing \cite{Doan11}. Due to the increasing popularity of mobile devices, mobile crowdsensing supports a broad range of sensing applications nowadays, ranging from social recommendation, such as restaurant recommendation, parking space discovery and indoor floor plan reconstrction \cite{Zhang17}, to environment monitoring, such as air quality measurement, noise level detection and dam water release warning.
With human intelligence and user mobility, mobile crowdsensing can significantly improve the trustworthiness of sensing data, extend the scale of sensing applications and reduce the cost on high-quality data collection \cite{Xiao17}.

While mobile crowdsensing makes data sensing appealing than ever, it also brings new challenges towards mobile users, one of which is privacy leakage, indicating that mobile crowdsensing puts the privacy of mobile users at stake \cite{Yang152,Krontiris14,Ni17}. The sensing data collected from the surrounding areas are necessarily people-centric and related to some aspects of mobile users and their social setting: where they are and where they are going; what places they are frequently visited and what they are seeing; how their health status is and which activity they prefer to do. Photos on social events may expose the social relations, locations or even political affiliations of mobile users. Furthermore, the more sensing tasks mobile users engaged in and the richer data the users contribute to, the higher probability that their sensitive information may be exposed with. Therefore, preserving the privacy of mobile users is the first-order security concern in mobile crowdsensing. If there is no effective privacy-preserving mechanism to protect the private information for mobile users, it is of difficulty to motivate mobile users to join in mobile crowdsensing services.
In addition, the sensing tasks may contain sensitive information about the customers who issue them. Some personal information about the customers, such as identities, locations, references and purchase intentions, can be predicted by curious entities from the releasing tasks. For example, a house agency may know Bob desire to buy a house in a particular area if Bob releases tasks to collect traffic condition and noise level in the neighborhood.
To preserve the privacy for both customers and mobile users, several privacy-preserving mobile crowdsensing schemes \cite{Cornelius08,Huang12,Dimitriou12,Qiu15} have been proposed by utilizing anonymity techniques. Nevertheless, anonymity is insufficient for privacy preservation, since the mobile users may be traced from travel routes and social relations. It is possible to uniquely identify 35\% of mobile users based on their top-two locations and 85\% of them from their top-three locations based on a large set of call data records provided by a US nationwide cell operator \cite{Zang11}. Therefore, it is important to explore strong privacy-preserving mechanisms to prevent privacy leakage for customers and mobile users in mobile crowdsensing.

Once all information about mobile users and customers is perfectly preserved, it is impossible for service providers to accurately recruit mobile users for task performing, while task allocation is a critical component in mobile crowdsensing to ensure the quality of sensing results. Different from traditional sensing networks, the produced data cannot be predicted as a priori, and their trustworthiness totally depends on the intelligence and behaviors of mobile users. In general, the higher quality the sensing data have, the more efforts and costs the mobile users should pay. Therefore, the set of mobile users would directly impact the quality of sensing data. How to identify the right groups of mobile users to produce the desired data according to the targets of sensing tasks is a complex problem from the service provider's perspective. Geography-based and reputation-based approaches are popular in mobile crowdsensing to allocate tasks to mobile users, but either has its inherent weaknesses. Firstly, reputation-based task allocation mechanisms \cite{Huang12,Ren15,Mousa15,Wang14} need a trusted third party (TTP) to perform heavy reputation management and are vulnerable to reputation-linking attacks, in which the anonymous mobile users can be re-identified from their reputations. Secondly, geography-based task allocation schemes can optimize users selection based on their spatial and temporal correlation \cite{Pournajaf14}, but it discloses the content of sensing tasks and the locations of mobile users to the service provider, while location privacy is one of the primary concerns for mobile users in pervasive environments. In summary, privacy preservation and task allocation become a pair of contradictory objectives in mobile crowdsensing.

 To resolve this issue, we propose a \underline{S}trong \underline{P}rivacy-preserving m\underline{O}bile cr\underline{O}wdse\underline{N}sing scheme (SPOON) supporting location-based task allocation, decentralized trust management and privacy preservation for both mobile users and customers simultaneously. By leveraging blind signatures and randomizable matrix multiplication, we fully prevent the privacy leakage from all sources for both mobile users and customers, including locations, identities and credit points, without scarifying the normal mobile crowdsensing services of service providers, such as task allocation, data filtering and trust management. The main contributions of this paper are summarized as three folds:

\begin{itemize}
  \item We design a privacy-preserving location matching mechanism based on matrix multiplication to allow service providers to allocate sensing tasks based on the sensing areas of tasks and the geographic locations of mobile users. Specifically, the service provider can determine whether a mobile user is in the sensing area of a task from two randomized matrices generated from the sensing area and the user's location. Thus, the service provider can learn the result of location matching, but has no knowledge about the interested areas of customers and the locations of mobile users.

  \item By extending the proxy re-encryption and BBS+ signature, we protect the sensitive information for mobile users and customers to prevent privacy leakage, including their identities, credit points, sensing tasks and sensing reports. Specifically, we allow the registered customers and mobile users to anonymously prove their capacities and trust levels to participate in the crowdsensing services and securely perform the sensing tasks without exposing contents of sensing tasks and sensing reports. Besides, to prevent the mobile users from misbehaving for unfair rewards, a trusted authority enables to detect the greedy mobile users and trace their identities.

  \item We introduce a privacy-preserving credit management mechanism for mobile users, in which mobile users are able to prove their trustworthiness without the exposure of credit points and the management of centralized servers. In particular, it supports the positive and negative updates of credit points for mobile users based on the contributions on the tasks. In addition, multiple service providers can cooperatively maintain a unique trust evaluation system, in the way that mobile users are allowed to participate in the mobile crowdsensing services offered by different service providers using unique credit points.
\end{itemize}

The remainder of this paper is organized as follows. We review the related work in section \ref{sec7}, and formalize system model, threat model and identify security goals in section \ref{Sec2}. In section \ref{sec3}, we propose our SPOON scheme, followed by the security discussion in section \ref{sec4}. In section \ref{sec5}, we discuss some extensions on SPOON and evaluate performance in section \ref{sec6}. Finally, we draw the conclusion in section \ref{sec8}.

\section{Related Work} \label{sec7}
Mobile crowdsensing has attracted great interests from the research community in recent years, especially for security and privacy aspects. To build a secure mobile crowdsensing architecture, AnonySense \cite{Cornelius08} was proposed to allow mobile devices to deliver sensing data through Mix networks. Christin et al. \cite{Christin11} investigated the location privacy of mobile users and presented a decentralized and collaborative mechanism to allow mobile users to exchange the sensing data when they physically meet for the protection of the travel routes of mobile users.
Dimitriou et al. \cite{Dimitriou12} raised the problem of customer's privacy leakage and designed a privacy-preserving access control scheme for mobile sensing to preserve the privacy for customers. However, none of above schemes enables to preserve the privacy for both mobile users and customers simultaneously. Therefore, Cristofaro and Soriente \cite{Cristofaro13} proposed a privacy-enhanced participatory sensing infrastructure (PEPSI) based on the blind extraction technique. In PEPSI, the identity-based encryption is extended to achieve the anonymity for both mobile users and customers, and a blind matching method is built to find the sensing reports for a specific task. Unfortunately, G\"unther et al. \cite{Guinther14} demonstrated PEPSI is vulnerable to collusion attacks across mobile users and customers. As a result, PEPSI fails to preserve the privacy of mobile users. To fix this drawback, a new infrastructure is designed from anonymous identity-based encryption. Qiu et al. \cite{Qiu15} presented SLICER, a $k$-anonymous privacy-preserving scheme for mobile sensing that achieves strong privacy preservation for mobile users and high data quality, by integrating a data coding technique and message transfer strategies. However, these schemes may be insufficient to preserve the privacy nowadays, since it is possible to re-identify the mobile users or customers through the combination of information from different sources, such as travel routes, social relations or payment records. To protect the sensing data, Zhou et al. \cite{Zhou16} introduced a generalized efficient batch cryptosystem to achieve both batch encryption and batch decryption from public key encryption, and extended to support fine-grained multi-receiver multi-file sharing in cloud-assisted mobile crowdsensing. Chen et al. \cite{Chen14} introduced a group management protocol to guarantee differential privacy of personal data to prevent the disclosure of sensing data.
Jin et al. \cite{Jin16} integrated user incentive, data aggregation and data perturbation mechanisms to design an incentivizing privacy-preserving data aggregation scheme to generate high-accurate aggregated results, and provide privacy protection for mobile users in mobile crowdsensing.

However, after the privacy of mobile users and customers is preserved, it is difficult for the service provider to find proper mobile users for task fulfillment.
Kazemi and Shahabi \cite{Kazemi12} focused on spatial task assignment for spatial crowdsourcing, in which the service provider allocates tasks based on the locations of mobile users. To hide their locations, To et al. \cite{To14} introduced a framework to protect the locations of mobile users based on differential privacy and geocasting. This framework provides heuristics and optimizations to determine effective geocast regions for reaching high task assignment ratio. Kazemi et al. \cite{Kazemi13} defined reputation scores to represent the probability that a mobile user can perform a task correctly, and a confidence level to state that a task is acceptable if its confidence is higher than a given threshold. Huang et al. \cite{Huang12} demonstrated that mobile users are vulnerable to linking attacks if they naively reveal their reputations to the service provider and presented an anonymization scheme from pseudonyms and a reputation management mechanism by employing a trusted server to minimize the risk of such attack.
Christin et al. \cite{Christin13} proposed an identity privacy-preserving reputation framework, which uses pseudonyms to preserve the identity privacy for mobile users. The pseudonyms cannot be linked in multiple time periods, while the reputations can be transferred for the mobile user that are associated with adjacent time periods. Consequently,
Wang et al. \cite{Wang14} proposed ARTSense to achieve the trust management without identity exposure in mobile sensing. ARTSense achieves both positive and negative updates of reputations for mobile users with no TTP, but it still requires a reputation database for each service provider to support reputation management. Moreover, the reputations of mobile users and the privacy of customers are directly revealed to the service provider and other curious entities in ARTSense.

For the above reasons, in our preliminary work \cite{NIICC17}, we proposed a privacy-preserving mobile crowdsensing framework to achieve trajectory-based task allocation without privacy leakage for both mobile users and customers. In this paper, we extend this work to support privacy-preserving decentralized credit management in mobile crowdsensing. In specific, the proposed SPOON (1) provides strong privacy preservation for mobile users; (2) protects the identities, sensing areas and tasks for customers; (3) allows the service provider to allocate sensing tasks based on the locations and credit points of mobile users; and (4) supports privacy-preserving credit management without a centralized server. We show the comparison on features between SPOON and the existing works in Table I.

\begin{table*}
\caption{Features Comparison on SPOON and other works}

\centering
\begin{tabular}{|c|c|c|c|c|c|c|c|c|c|}
\hline
\multirow{2}{*}{Features} &\multicolumn{2}{|c|}{Identity Privacy} & \multicolumn{2}{|c|}{Data Privacy} & \multicolumn{2}{|c|}{Location Privacy}  & \multicolumn{3}{|c|}{Credit Management}  \\
\cline{2-10}
& Users& Customers & Users& Customers& Users& Customers&Credit Privacy \& Sharing & No TTP &Greedy user Tracing\\
\hline\hline
SPOON&$\surd$&$\surd$&$\surd$&$\surd$&$\surd$&$\surd$&$\surd$&$\surd$&$\surd$ \\
\hline
\cite{Cornelius08,Huang12} &$\surd$&X&X&X&X&X&X&X&X\\
\hline
\cite{Dimitriou12}&X&$\surd$&X&$\surd$&X&$\surd$&X&X&X\\
\hline
\cite{Qiu15,Kazemi12,Christin13}&$\surd$&X&$\surd$&X&$\surd$&X&X&X&X\\
\hline
\cite{Wang14} &$\surd$&X&$\surd$&X&$\surd$&X&X&$\surd$&$\surd$\\
\hline
\cite{Christin11} &X&X&X&X&$\surd$&X&X&X&X\\
\hline
\cite{Cristofaro13,Guinther14} &$\surd$&$\surd$&$\surd$&X&X&X&X&X&X\\
\hline
\cite{Zhou16,Chen14,Jin16} &X&X&$\surd$&X&X&X&X&X&X\\
\hline
\end{tabular}
\end{table*}

\section{Problem Statement} \label{Sec2}
In this section, we formally define the system model and threat model, and identify our design goals.
\subsection{System model}
The mobile crowdsensing service provides customers a people-centric way for data collection from surrounding environment. The architecture consists of three entities: a service provider, customers and mobile users, as shown in Fig. 1.

\begin{figure}
\begin{center}
\centerline{\includegraphics[width=0.5\textwidth]{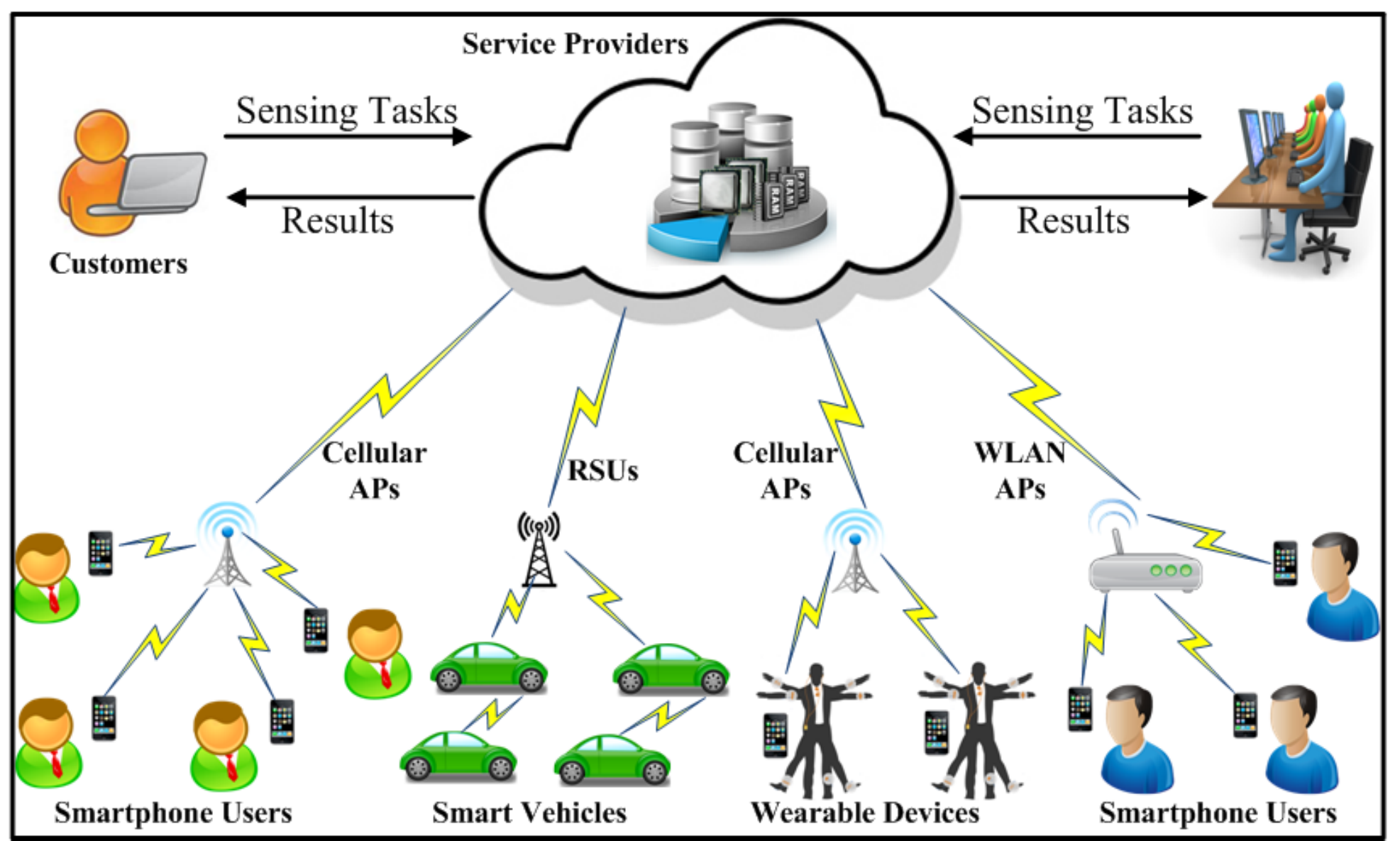}}
\caption{Mobile Crowdsensing Architecture.}
\label{fig:two}
\end{center}
\end{figure}

{\sf Service Providers}: Service providers develop cloud services by themselves or rent the cloud resources offered by cloud service providers. They have sufficient storage and computing resources to provide mobile crowdsensing services. The service providers receive sensing tasks from customers and allocate them to mobile users based on their locations. They collect sensing reports from mobile users, select sensing reports based on the credit points of mobile users and generate sensing results for customers. The service provider also distributes credit points to mobile users for incentive.

{\sf Customers}: The customers can be individuals, corporations or organizations. They need to accomplish data collection tasks, e.g., to study traffic congestion in a city, pollution level of a creek and satisfactory on public transportation, but they do not have sufficient capabilities to perform tasks by themselves. Thereby, they issue their sensing tasks to the service providers to obtain the sensing results.


{\sf Mobile Users}: Every mobile user has several mobile devices, e.g., mobile phones, tablets, vehicles and smart glasses. These mobile devices, with rich computational, communication and storage resources, are carried by their owners wherever they go and whatever they do. The mobile users make sure the battery on mobile devices have sufficient power to support their normal functions. The mobile users participate in sensing tasks and utilize their portable devices to collect data from their surrounding areas to fulfill sensing tasks, and report sensing data to the service providers for earning credit points.

\subsection{Threat Model}
The service provider is responsible for offering mobile crowdsensing service to customers, but it may strive to increase the income and violate its privacy policy of data protection. For example, Uber, a crowdsourcing-based ride-sharing service provider, made ride-booking data publicly accessible without the permission of customers in January, 2017, for its own purpose. Therefore, the service provider is not fully trusted, but honest-but-curious.
On one hand, the service provider would honestly perform the mobile crowdsensing service; one the other hand, it may learn a spatio-temporal probability distribution for a specific mobile user and other sensitive information about customers and mobile users, e.g., preference, social relation, political affiliation and purchase intention, from the maintained information, including sensing tasks and sensing reports. Moreover, the employees in service provider may capture and exploit the sensitive information about mobile users.

Mobile users are interested in the privacy about the customers and the other mobile users. In particular, they are willing to know the other mobile users participating in the same tasks, and learn more information about customers they are working for to reach the expectations of customers. Further, mobile users may be greedy for the credit points, such that they may anonymously submit more sensing reports than allowed to warn unfair credit points. In addition, the mobile users may maliciously forge, modify the sensing data or deliver ambiguous, biased sensing data to cheat customers for credit points. These forged or biased data can be discovered using redundancy or truth discovery approaches. The locations are extracted from GPS trusted chips in mobile devices or access points, we assume that mobile users cannot modify their location information.

The external attackers, such as eavesdroppers and hackers, also bring serious security threats towards mobile crowdsensing services. It is possible for an attacker to obtain the identities of the nearby mobile users or customers via physical observation, such that the anonymity may be insufficient for privacy preservation for customers and mobile users. The customers are fully trusted since they are the main beneficiaries of mobile crowdsensing service.

\subsection{Design Goals} \label{security}
To enable strong privacy-preserving mobile crowdsensing under the
aforementioned system model and against security threats, SPOON should achieve the following design goals:

\begin{itemize}
  \item  {\sf Location-based Task Allocation}: The sensing tasks are allocated to the mobile users in the sensing areas defined by the customers, and other mobile users out of the given areas cannot learn any information about the tasks.
  \item {\sf Location Privacy Preservation}: The locations of mobile users and the sensing areas of sensing tasks would not be exposed to others. The mobile users are only aware whether they are in the sensing area or not.
  \item {Data Confidentiality}: No entity, except the delegated participants, can obtain the content of releasing tasks or sensing reports, such that the privacy of customers and mobile users would not be disclosed to others.
  \item {\sf Anonymity of Mobile Users and Customers}: The customers, mobile users, the service provider or their collusion are unable to link a sensing report to a mobile user or link a sensing task to a customer. It is even impossible for an attacker to identify whether two sensing reports are generated by the same mobile user or two sensing tasks are issued by the same customer.
  \item {\sf Privacy-Preserving Credit Management}: Credit points are used to represent the reputation of mobile users and encourage them to participate in the mobile crowdsensing activities as rewards. The service provider selects the sensing reports based on the credit points of mobile users and awards credit points to mobile users without knowing the exact credit points of mobile users. The balance of credit points is achieved, which means that it is impossible for the mobile users to forge credit points without being detected, such that the total credit points of a mobile user should be equal to the awarded credit points plus the initial points.
  \item{\sf Greedy User Tracing}: The identities of greedy mobile users, who submit more than one sensing report for the same task in a reporting period, are recovered to prevent the mobile user from awarding unfair credit points.
\end{itemize}

\section{The SPOON Scheme} \label{sec3}
 In this section, we review the preliminaries and propose our SPOON, which is composed of five phases, {Service Setup},
{User Registration}, {Task Allocation}, {Data Reporting} and {Credit Assignment}, based on the matrix multiplication, the BBS+ signature \cite{Au06} and the proxy re-encryption \cite{Ateniese05}.

\subsection{Preliminaries}
We review the preliminaries that are used to design our SPOON, including the bilinear map, the BBS+ signature and the proxy re-encryption.

{\sf Bilinear Map}. Let $(\mathbb{G}, \mathbb{G}_T)$ be two cyclic groups with a prime order $p$. $\hat{e}: \mathbb{G} \times \mathbb{G}\rightarrow \mathbb{G}_T$ is the bilinear map with the following properties:
\begin{itemize}
  \item Bilinearity. For $g \in \mathbb{G}$, $a, b \in \mathbb{Z}_p$, $\hat{e}(g^a,g^b)=\hat{e}(g,g)^{ab}$.
  \item Non-degeneracy. For $g \neq 1_{\mathbb{G}}$, $\hat{e}(g,g)\neq 1_{\mathbb{G}_T}$.
  \item Computability. $\hat{e}$ is efficiently computable.
  \item (Unique Representation). The binary presentation for all elements in $\mathbb{G},\mathbb{G}_T$ is unique.
\end{itemize}

{\sf BBS+ Signature \cite{Au06}}. Here we briefly review the BBS+ signature due to \cite{Au06}, which can be utilized to sign $\ell$-message vector $(m_1,\cdots, m_{\ell})$.

Let $g, g_1, \cdots, g_{\ell+1}$ be generators of $\mathbb{G}$. Randomly choose $x$ from $ \mathbb{Z}_p$ as the secret key of the signature scheme, and compute the corresponding public key as $y=g^x$.

A signature on messages $(m_1, \cdots, m_{\ell})$ is $(A, e,s)$, where $A=(g g_1^{m_1} \cdots g_{\ell}^{m_{\ell}} g_{\ell+1}^{s})^{\frac{1}{x+e}}$ and $(e,s)$ are random values chosen from $\mathbb{Z}_p$.

This signature can be checked as: $\hat{e}(g g_1^{m_1}\cdots g_{\ell}^{m_{\ell}} g_{\ell+1}^{s}, g) \stackrel{?}{=} \hat{e}(A, yg^e)$.

The security of BBS+ signature can be reduced to the $q$-SDH assumption and it can be utilized to construct a zero-knowledge proof-of-knowledge protocol that allows the signer to prove the possession of the message-signature pair.\\

{\sf Proxy Re-Encryption \cite{Ateniese05}}. Proxy Re-encryption is a special public key encryption with a desirable property that a semi-trusted proxy enables to convert a ciphertext for Alice into a ciphertext for Bob without seeing the underlying plaintext, given a proxy re-encryption key. Thanks to this promising property, it has been widely employed in data sharing scenarios. The proxy re-encryption scheme is proposed by Ateniese et al. \cite{Ateniese05}, the details of which are as follows:

\begin{itemize}
\item KeyGen$(\cdot)$. Alice picks a random value $a \in \mathbb{Z}_p$ as the secret key $sk_a$ and compute the public key $pk_a=g^a$.
  \item RKeyGen$(sk_a, pk_b)$. Alice delegates to Bob by sending the re-encryption key $rk_{A\rightarrow B}=g^{b/a}$ to a proxy by using Bob's public key.
  \item Encrypt$(m, pk_a)$. To encrypt a message $m \in \mathbb{G}_T$ under $pk_a$, Alice chooses a random value $k \in \mathbb{Z}_p$ to compute $c_a=(g^{ak}, m\hat{e}(g,g)^{k})$.
  \item Re-Enc$(c_a, rk_{A\rightarrow B})$. The proxy can change the ciphertext $c_a$ into a ciphertext $c_b$ for Bob with $rk_{A\rightarrow B}$. From $c_a$, the proxy calculates $\hat{e}(g^{ak},g^{b/a})=\hat{e}(g,g)^{bk}$ and releases $c_b=(\hat{e}(g,g)^{bk}, m\hat{e}(g,g)^{k})$.
  \item Decrypt $(c_b, sk_b)$. Bob enables to decrypt $c_b$ to obtain $m$ as $m=m\hat{e}(g,g)^{k}/(\hat{e}(g,g)^{bk})^{1/b}$.
\end{itemize}

\begin{table}
\caption{Frequently Used Notions}\label{tab:one}
\vspace{-0.25in}
\begin{center}
\begin{tabular}{|l|l|}
\hline
${U_i}_{\{i \in R\}}$   & Set of registered mobile users \\\hline
${U_i}_{\{i \in \mathcal{L}\}}$    & Set of mobile users in sensing area $L$ \\\hline
$ST$            & A task issued by a customer\\\hline
$task$     & The detailed content of a task $ST$ \\\hline
$expires$     & The expiration time of a task $ST$  \\\hline
$area$     & The sensing region of a task $ST$ \\\hline
$L_{m\times n}$    & A matrix to represent the service area of the service provider\\\hline
$\widehat{L}_{m\times n}$  & A matrix to represent the sensing area of a task $ST$\\\hline
$\widetilde{L}_{m\times n}$       & A matrix to represent the current and future locations of a user  \\\hline
$\widehat{M}_{m\times n}$    & A random invertible matrix\\\hline
$\widetilde{M}_{m\times n} $    & A random invertible matrix\\\hline
$I$ & The unique identity of a registrant (mobile user or customer)\\\hline
$P_0$     & The initial credit point of a mobile user \\\hline
$\epsilon$     & The trust level of a sensing report \\\hline
$\gamma$     & The maximum of trust level in a task $ST$\\\hline
$Q$     & The credit threshold chosen by a mobile user \\\hline
$A,e,s$     & The anonymous credential of a mobile user or customer \\\hline
$B,f,t$     & The anonymous credential of a mobile user with credit point $P$ \\\hline
\end{tabular}
\end{center}
\end{table}%

\subsection{High-Level Description}
We first provide a high-level description of SPOON and its information flow, which is shown in Fig. 2. The notions frequently used in SPOON are listed in Table II.

{\sf Service Setup}: A trusted authority (TA) bootstraps the whole mobile crowdsensing service for the service provider by defining the public parameters $(\mathbb{G}, \mathbb{G}_T, p, g,g_0,g_1,g_2,g_3,h,h_0,h_1,h_2,h_3,h_4, G,H,\mathcal{G},\mathcal{H},\mathcal{F})$ and generates its secret-public key pair $(\alpha, T)$. The service provider also generates the secret-public key pair $(\beta,S)$, and defines a matrix $L_{m\times n}$ to denote the geographic region of its crowdsensing service.

{\sf User Registration}: The TA registers the mobile users and customers, who are willing to participate in the mobile crowdsensing service. It evaluates the registrant to determine the initial credit point $P_0$ and interacts with the registrant to generate an anonymous credential $(A,e,s,B,f,t)$. $(A,e,s)$ is used to access the mobile crowdsensing service and $(B,f,t)$ is used to credit management for the registrant. To achieve the anonymity, the ownership of $(A,e,s)$ and $(B,f,t)$ is proved by the registrant for identity authentication and credit evaluation using zero-knowledge proofs, respectively. Besides, $RK$ is assigned to the registrant for the decryption of allocated sensing tasks.

{\sf Task Allocation}: A customer generates a sensing task $ST$ and sends the message $(c_1,c_2,c_3,expires,\widehat{N}_{n \times n}, \gamma,w,\mathcal{PK}_2)$ to the service provider, which consists of the encrypted task $(c_1,c_2,c_3)$, the expiration time $expires$, the randomized sensing area $\widehat{N}_{n \times n}$, the identity proof $\mathcal{PK}_2$ and other information. The latter releases $(num, expires,\gamma)$ to attract mobile users for participation, where $num$ is the identifier of $ST$. A mobile user $U_{i \{i\in R\}}$ sends its location $\widetilde{N}_{n \times n}$ and identity proof $\mathcal{PK}_3$ to the service provider. Then, the service provider finds the set of mobile users ${U_i}_{\{i \in \mathcal{L}\}}$  in the sensing area of $ST$ based on two matrices ($\widehat{N}_{n \times n}, \widetilde{N}_{n \times n}$). Since ($\widehat{N}_{n \times n}, \widetilde{N}_{n \times n}$) are randomized matrices, the service provider can learn whether $U_i$ is in the sensing area of $ST$ based on matrix multiplication, but has no information about $ST$'s sensing area and $U_i$'s location. The service provider re-encrypts the ciphertext $(c_1,c_2,c_3)$ to be decryptable for ${U_i}_{\{i \in \mathcal{L}\}}$ using $\beta$. Finally, the service provider sends $(num,c_2,c_3,c_4, expires,\gamma,w)$ to ${U_i}_{\{i \in \mathcal{L}\}}$.

{\sf Data Reporting}: ${U_i}_{\{i \in \mathcal{L}\}}$ encrypts the collected data $m_i$ to generate $(D_i,D'_i)$, and sends the sensing report $(num, D_i, D'_i, C'_i,X_i,Y_i,Z_i,Q_i,\tau_j, \mathcal{SPK})$ to the service provider, in which $C'_i$ is the commitment on the identity $I_i$ and credit point $P_i$, $X_i$ is the identifier of this report, $Y_i$ is the identifier of $U_i$, $Z_i$ is a tag to identify the double-reporting user, $Q_i$ is the claimed credit threshold to show that the number of credit points $U_i$ has is larger than $Q_i$, $\tau_j$ is the current slot for reporting, and $\mathcal{SPK}$ is used to prove the ownership of its credit points $P_i$. The service provider selects $w-$sensing reports based on the claimed thresholds and forwards the selected reports to the customer. The TA can recover the identity of anonymous mobile user who double-reports sensing reports with the service provider using the double-reporting tag $Z_i$.

{\sf Credit Assignment}: The customer evaluates the trustworthiness of each report and returns the corresponding trust level $\epsilon_i \in [-\gamma,\gamma]$ to the service provider. The latter computes the number of credit points awarded to $U_i$, $\theta_i$, and forwards $(B_i,t''_i,f_i,\theta_i,Y_i)$ to $U_i$, where $(B_i,t''_i,f_i)$ is the ticket for awarded credit points $\theta_i$, and $Y_i$ is used to identify the mobile user $U_i$. Once receiving $(B_i,t''_i,f_i,\theta_i,Y_i)$, $U_i$ updates its credit points $P'_i=P_i+\theta_i$ and the anonymous credential $(B_i,f_i,t_i)$ for the new $P'_i$.

\begin{figure}
\begin{center}
\centerline{\includegraphics[width=0.5\textwidth]{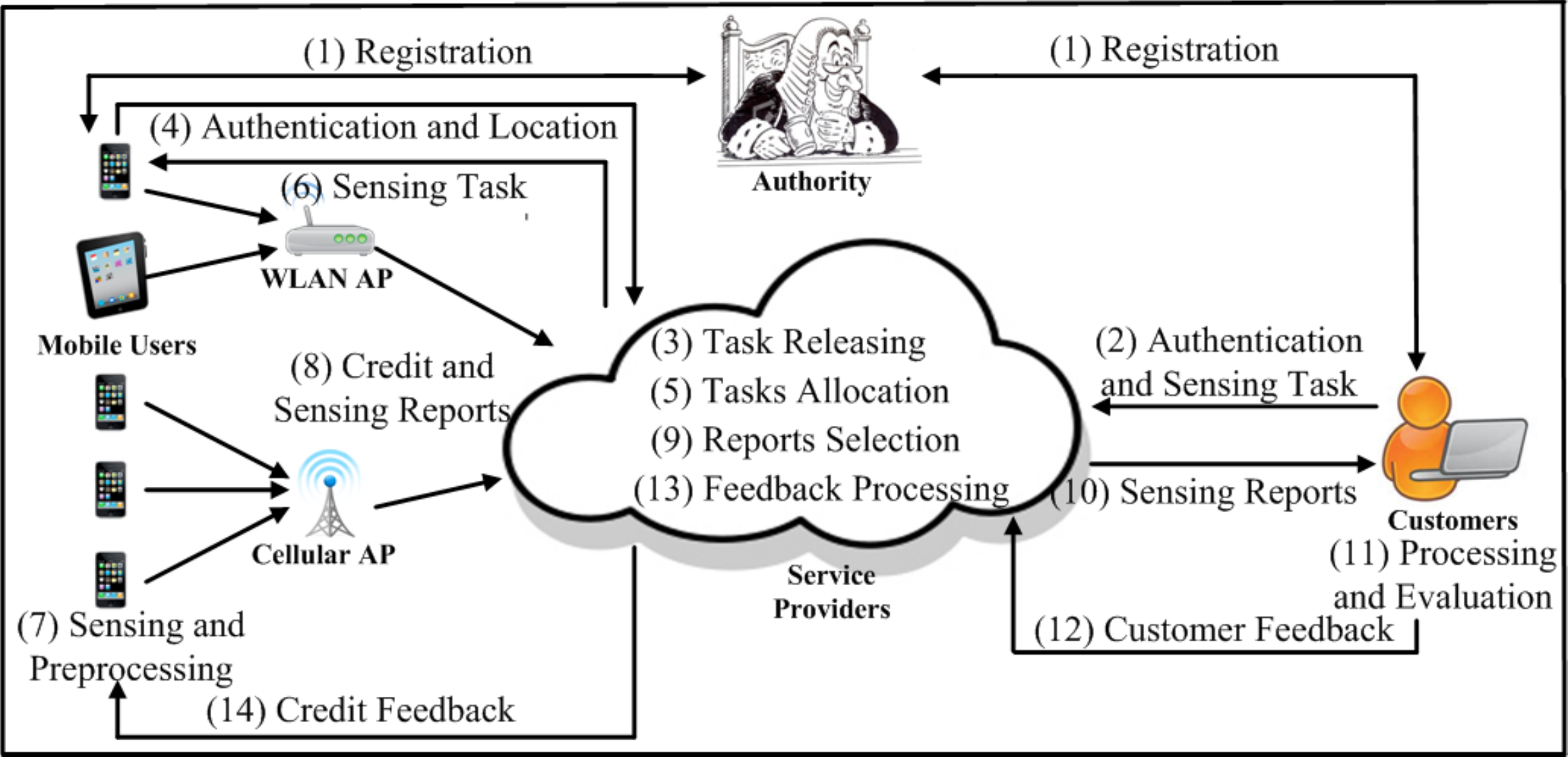}}
\caption{Information Flow of Mobile Crowdsensing.}
\label{fig:two}
\end{center}
\end{figure}

\subsection{The Detailed SPOON}
We then show the detailed SPOON as follows.
\subsubsection{{\sf Service Setup}}
Let $(\mathbb{G}, \mathbb{G}_T)$ be two cyclic groups with a prime order $p$, where $p$ is $\lambda$ bits, and $\hat{e}: \mathbb{G} \times \mathbb{G}\rightarrow \mathbb{G}_T$ be a bilinear map. The authority picks random generators $g,g_0,g_1,g_2,g_3, h,h_0,h_1,h_2,h_3,h_4 \in \mathbb{G}$ and computes $G=\hat{e}(g,g)$ and $H=\hat{e}(h,h)$ respectively. The TA also chooses a random value $\mathcal{G} \in \mathbb{G}_T$ and defines a cryptographic hash function $\mathcal{H}:\{0,1\}^* \rightarrow \mathbb{Z}_p$ and a pseudo-random function $\mathcal{F}: \mathbb{Z}_p \times \{0,1\}^* \rightarrow \mathbb{Z}_p$. The public parameters $param$ are $(\mathbb{G}, \mathbb{G}_T, p, g,g_0,g_1,g_2,g_3,h,h_0,h_1,h_2,h_3,h_4, G,H,\mathcal{G},\mathcal{H},\mathcal{F})$. The TA randomly chooses $\alpha \in \mathbb{Z}_p$ as its secret key and calculates the public key $T=g^{\alpha}$.

To setup the mobile crowdsensing service, the service provider randomly chooses its secret key $\beta \in \mathbb{Z}_p$ and computes $S=h^{\beta}$ as its public key. It also employs a matrix $L_{m \times n}$ to denote the geographical region that the crowdsensing service can cover according to the longitude and latitude. Each entry in the matrix denotes a small grid in the sensing region, as shown in Fig. 3. Assume the longitude of Ontario is from $74.40^{\circ}$W to $95.15^{\circ}$W, the latitude is from $41.66^{\circ}$N to $57.00^{\circ}$N, we can use a $208 \times 154$ matrix or $2075 \times 1534$ matrix more precisely to represent the Ontario region.

\subsubsection{{\sf User Registration}}
Either customer or mobile user is required to register at the TA to obtain an anonymous credential, which is used to participate in the crowdsensing service. Each registrant is assigned a unique identity $I$ in the system, which can be the telephone number or mailing address in practise. The registrant picks three random values $s',a,t' \in \mathbb{Z}_p$ to compute $C=g_1^{s'}g_2^a$, $C'=h_1^{t'}h_2^a$, $\widehat{A}=g^a$, and sends $(I,C,C',\widehat{A})$ to the TA, along with the following zero-knowledge proof: $$\mathcal{PK}_1\{(s',t',a): C=g_1^{s'}g_2^a \land C'=h_1^{t'}h_2^a\land \widehat{A}=g^a\}.$$
The TA firstly checks the proof $\mathcal{PK}_1$ for ensuring that $(C,C',\widehat{A})$ are generated correctly. Then, it evaluates the registrant's initial credit point according to its credit record, which is assumed to be $P_0$. After that, the TA randomly picks $s'',e,t'',f \in \mathbb{Z}_p$ to calculate $A=(g_0Cg_1^{s''}g_3^I)^{\frac{1}{\alpha+e}}$, $B=(h_0C'h_1^{t''}h_3^I h_4^{P_0})^{\frac{1}{\alpha+f}}$, $RK=\widehat{A}^{\frac{1}{\alpha}}$, and returns $(A,B, s'',t'', e, f,P_0, RK)$ to the registrant through secure channels. Finally, the TA stores the tuple $(I,P_0,\widehat{A})$ in its database.

The registrant computes $s=s'+s'', t=t'+t''$ and checks
\begin{center}
$\hat{e}(A,Tg^e)\stackrel{?}{=}\hat{e}(g_0g_1^sg_2^ag_3^I,g)$, ~~~~ $\hat{e}(B,Th^f)\stackrel{?}{=}\hat{e}(h_0h_1^th_2^ah_3^Ih_4^{P_0},h)$.
\end{center}
The registrant stores $(A,e,s,B,f,t,a,I,P_0,\widehat{A}, RK)$ secretly on the read-only memory of mobile device.

\subsubsection{{\sf Task Allocation}}
A customer with registered information $(A,e,s,B,f,t,a,I,P_0,\widehat{A}, RK)$ has a sensing task to be allocated to mobile users and requests the sensing data slot by slot, where each slot ranges from minutes to days depending on the specific requirements of the sensing task. The statement of the task is defined as $ST=(task, expires, area, \gamma,w)$, which indicate the content (what to sense), the expiration time (when to sense), the sensing area (where to sense), the maximum trust level and the number of required reports, respectively. Other attributes (e.g., sensing intervals, acceptance conditions, benefits, reporting periods) can be illustrated in $task$. To protect the content of the task, the customer randomly picks $k,r_1,r_2,r_3 \in \mathbb{Z}_p$ to calculate $u=g^k, c_1=S^{r_2}$, $c_2=T^{r_1}$ and $c_3=(task||u)G^{r_1}H^{r_2}$. Then, the customer generates a matrix $\widehat{L}_{m \times n}$ to indicate the target sensing region $area$. As depicted in Fig. 3, for each position in the sensing area, the corresponding entry in $\widehat{L}_{m \times n}$ is set to be a random value chosen from $\mathbb{Z}_p^*$, and the value for a location outside is set to be zero. To mask the sensing area in $\widehat{L}_{m \times n}$, the customer picks $m \times n$ random numbers from $\mathbb{Z}_p^*$ to generate an invertible matrix $\widehat{M}_{m \times n}$ and computes $\widehat{N}_{n \times n}=\widehat{L}_{m \times n}^T \cdot \widehat{M}_{m \times n}$, where $\widehat{L}_{m \times n}^T$ is the transpose of the matrix $\widehat{L}_{m \times n}$. Note that all non-zero entries in $\widehat{L}_{m \times n}$ should be distinct, unless an attacker still can learn the sensing region from $\widehat{N}_{n \times n}$. Finally, the customer keeps $k$ in private and sends $(c_1,c_2,c_3,expires,\widehat{N}_{n \times n}, \gamma,w)$ to the service provider, along with the following zero-knowledge proof: $$\mathcal{PK}_2\{(A,e,s,a, I): \hat{e}(A,Tg^e){=}\hat{e}(g_0g_1^sg_2^ag_3^I,g)\}.$$

\begin{figure}
\centerline{\includegraphics[width=0.3\textwidth]{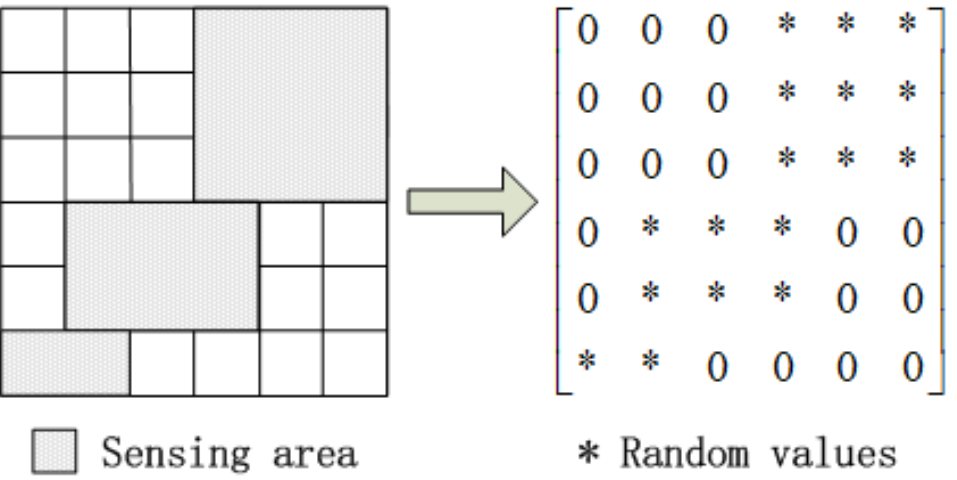}}
\caption{Sensing Area and the Matrix $\widehat{L}_{6 \times 6}$.}
\label{fig:three}
\end{figure}

The service provider checks the validity of the proof $\mathcal{PK}_2$. If yes, it assigns a task identifier $num$, releases $(num, expires,\gamma)$ and stores $(num, c_1,c_2,c_3,expires,\widehat{N}_{n \times n},\gamma,w)$ in its database.

When a mobile user ${U_i}_{\{i \in R\}}$ with $(A_i,e_i,s_i,B_i,f_i,$ $t_i,a_i,I_i,P_i,\widehat{A}_i, RK_i)$ is willing to participate in crowdsensing activities, it firstly picks a random value $\nu \in \mathbb{Z}_p$ to calculate $\mu=h^{\nu}$. Then, ${U_i}$ generates a matrix $\widetilde{L}_{m\times n}$ according to its current location and the places it will visit. For each location $U_i$ will reach, the corresponding entry in $\widetilde{L}_{m\times n}$ is set to be a random value chosen from $\mathbb{Z}_p^*$, and the rest entries are set to be zero. The non-zero entries in $\widetilde{L}_{m \times n}$ should be different. To protect these location information, it also generates a random invertible matrix $\widetilde{M}_{m\times n}$ by picking ${m\times n}$ random values from $\mathbb{Z}_p^*$, and calculates $\widetilde{N}_{n \times n}=\widetilde{M}_{m \times n}^T\cdot \widetilde{L}_{m \times n}$. Finally, ${U_i}$ keeps $\nu$ secretly and sends $(\mu, \widetilde{N}_{n \times n})$ to the service provider, along with the following zero-knowledge proof: $$\mathcal{PK}_3\{(A_i,e_i,s_i,a_i, I_i): \hat{e}(A_i,Tg^{e_i}){=}\hat{e}(g_0g_1^{s_i}g_2^{a_i}g_3^{I_i},g)\}.$$

The service provider returns failure if $\mathcal{PK}_3$ is invalid. Otherwise, for each unexpired task, it uses $\widehat{N}_{n \times n}$ to calculate $N_{n \times n}=\widetilde{N}_{n \times n} \cdot \widehat{N}_{n \times n}$ and checks whether $N_{n \times n}$ is zero matrix or not. If $N_{n \times n}$ is non-zero matrix, which means that $U_i$ can match $ST$, the service provider calculates $c_4=\hat{e}(\mu, c_1)^{\frac{1}{\beta}}$ and releases $(num,c_2,c_3,c_4, expires,\gamma,w)$ for $U_i$. If there is no task to match $U_i$, the service provider responds failure.

When $U_i$ obtains $(num, c_2,c_3,c_4, expires,\gamma,w)$, it decrypts $(c_2,c_3,c_4)$ by using $(\nu, a_i)$ as $task||u=c_3c_4^{-{\frac{1}{\nu}}}\hat{e}(c_2,RK_i)^{-{\frac{1}{a_i}}}$. Then, $U_i$ evaluates the task and determines to participate in or abandon this task according to benefit and cost. If $U_i$ accepts the task $ST$, it starts to perform the sensing work according to the details in $task$. The correctness of $task||u$ is elaborated as follows:
\begin{eqnarray}
& &c_3c_4^{-{\frac{1}{\nu}}}\hat{e}(c_2,RK_i)^{-{\frac{1}{a_i}}}\\
&=&c_3\hat{e}(\mu,c_1)^{-{\frac{1}{\beta \nu}}}\hat{e}(c_2,RK_i)^{-{\frac{1}{a_i}}} \nonumber\\
&=&(task||u)G^{r_1}H^{r_2}\hat{e}(h^{\nu},S^{r_2})^{-{\frac{1}{\beta \nu}}}\hat{e}(T^{r_1},g^{\frac{a_i}{\alpha}})^{-{\frac{1}{a_i}}} \nonumber\\
&=& (task||u)G^{r_1} H^{r_2}H^{-r_2}G^{-r_1} \nonumber\\
&=& task||u. \nonumber
\end{eqnarray}

\subsubsection{{\sf Data Reporting}}
$U_i$ collects and, pre-processes the data $m_i \in \mathbb{G}_T$ and submits a sensing report to the customer periodically, which includes the collection time, the sensing location and the detailed content. The reporting periods are defined by the customer, and we assume the current slot is $\tau_j$. To prevent attackers from learning $m_i$, $U_i$ uses $u$ to encrypt $m_i$ as $D_i=u^{\hat{r}_i}$, $D'_i=m_i G^{\hat{r}_i}$, where $\hat{r}_i$ is a value randomly chosen from $\mathbb{Z}_p$. Then, $U_i$ randomly picks $t'_i \in \mathbb{Z}_p$ to compute $C'_i=h_1^{t'_i}h_2^{a_i}h_3^{I_i}h_4^{P_i}$. Next, $U_i$ computes $X_i=\mathcal{H}(num||m_i||\tau_j)$, $v_i=\mathcal{F}_{a_i}(num||I||\tau_j)$, $Y_i=H^{v_i}$ and $Z_i=\hat{e}(g,\widehat{A}_i)\mathcal{G}^{X_iv_i}$. Finally, $U_i$ chooses a credit threshold $Q_i$ and sends the report $(num, D_i, D'_i, C'_i,X_i,Y_i,Z_i,Q_i,\tau_j)$ to the service provider, along with the following zero-knowledge proof:
$$\mathcal{SPK} \left \{ \begin{array}{c}  (B_i,f_i,t_i,t'_i, a_i,I_i,P_i,v_i):~~~~~~~~~~~~~~~~~~~ \\ ~~~~~~~~\hat{e}(B_i,Th^{f_i}){=}\hat{e}(h_0h_1^{t_i}h_2^{a_i}h_3^{I_i}h_4^{P_i},h) \land \\ ~~~~~~~~C'_i=h_1^{t'_i}h_2^{a_i}h_3^{I_i}h_4^{P_i}\land\\
~~~~P_i>Q_i \land\\
~~~~~~Y_i=H^{v_i}\land \\
~~~~~~~~Z_i=\hat{e}(g,\widehat{A}_i)\mathcal{G}^{X_iv_i} \end{array}\right \}(num).$$

The service provider returns failure if $\mathcal{SPK}$ is invalid; otherwise, the service provider checks whether there is another report $(num, \widetilde{D}_i, \widetilde{D}'_i, \widetilde{C}'_i,\widetilde{X}_i,Y_i,\widetilde{Z}_i,\widetilde{Q}_i)$ that has the same $Y_i$ and different $\widetilde{X}_i$ with the new received report $(num, D_i, D'_i, C'_i,X_i,Y_i,Z_i,Q_i)$. If yes, the service provider computes and sends $W=(\frac{\widetilde{Z}_i^{X_i}}{Z_i^{\widetilde{X}_i}})^{\frac{1}{X_i-\widetilde{X}_i}}$ to the TA, and the TA can find the mobile user's identity $I_i$ by utilizing $\widehat{A}_i$ in the database to check $W=\hat{e}(g,\widehat{A}_i)$. Such that, the identity of the greedy mobile user is recovered by the TA if it submits two different sensing reports in a reporting slot. Then, according to the claimed thresholds, the service provider chooses $w$ reports that have top-$w$ thresholds, and releases them for the customer. Note that the mobile users, whose reports are not selected, can increase their thresholds in the next reporting slot $\tau_{j}+1$.

When the customer retrieves the reports, it can decrypt them using the stored $k$ as $m_i=D'_i\hat{e}(g,D_i)^{\frac{1}{k}}$ one by one.

\subsubsection{{\sf Credit Assignment}}
After the customer obtains the sensing result, it evaluates the trustworthiness of each report and responds the corresponding trust level to the service provider. The trust level of $m_i$ is defined as $\epsilon_i \in [-\gamma,\gamma]$. If $\epsilon_i$ is positive, $m_i$ is trustworthy, otherwise, $m_i$ is incredible.

Upon receiving trust levels, the service provider randomly picks $t''_i,f_i \in \mathbb{Z}_p$ to compute $\theta_i=INT(\epsilon_i Q_i)$, $B_i=(h_0h_1^{t''_i}C'_ih_4^{\theta_i})^{\frac{1}{\beta+f_i}}$, and releases $(B_i,t''_i,f_i,\theta_i,Y_i)$ for $U_i$, where $INT(x)$ is the nearest integer function.

$U_i$ retrieves $(B_i,t''_i,f_i,\theta_i,Y_i)$ from the service provider, computes $t_i=t'_i+t''_i$, $P'_i=P_i+\theta_i$ and checks whether $\hat{e}(B_i,Sh^{f_i}){=}\hat{e}(h_0h_1^{t_i}h_2^{a_i}h_3^{I_i}h_4^{P'_i},h)$ or not. If yes, $U_i$ uses the new tuple $(B_i,f_i,t_i,P'_i)$ to replace the previous one and stores them together with $(A_i,e_i,s_i, a_i,T_i,\widehat{A}_i, RK_i)$. Meanwhile, $U_i$ updates $P'_i$ in the read-only memory, which can be used to show the credit points in the future crowdsensing activities.
Further, since $(B_i,f_i,t_i,P'_i)$ are managed by $U_i$, $U_i$ enables to prove the ownership of $(B_i,f_i,t_i)$ cross service providers. The credit points awarded by different service providers can be accumulated and $U_i$ can prove the credit points to multiple service providers during the participation of mobile crowdsensing services offered by different service providers.

\section{Security Discussion} \label{sec4}
In this section, we show that SPOON satisfies five security goals defined in \ref{security}: location privacy, anonymity, data confidentiality, credit balance and greedy user tracing.

\subsection{Location Privacy}
The sensing region of a task is represented as a matrix $\widehat{L}_{m\times n}$, which is randomized by a random matrix $\widehat{M}_{m\times n}$ to generate $\widehat{N}_{n\times n}$. The location of the mobile user is transformed to be $\widetilde{N}_{n\times n}$. Having two matrices $\widehat{N}_{n\times n}$ and $\widetilde{N}_{n\times n}$, the service provider cannot learn any information about the location of the mobile user and the sensing area of the task. The service provider computes ${N}_{n\times n}=\widehat{N}_{n\times n} \cdot \widetilde{N}_{n\times n}$. If there is no overlapping between the sensing area of task and the location of mobile user, $N_{n \times n}$ must be zero matrix. If one overlapping grid exists, whose corresponding entry is $\widehat{L}_{ij}$ in $\widehat{L}_{m\times n}$ and is $\widetilde{L}_{ij}$ in $\widetilde{L}_{m\times n}$, respectively, the entries in $j$-row of $\widehat{N}_{n\times n}$ are nonzero, as well as the entries in $j$-column of $\widetilde{N}_{n\times n}$. Thus, the service provider enables to know that there are some overlapping locations on the $j$-column of the sensing area, while it is unable to distinguish which location is overlapped from $m$ locations. Further, $\widehat{N}_{n\times n} \cdot \widetilde{N}_{n\times n}$ and $\widetilde{N}_{n\times n} \cdot \widehat{N}_{n\times n} $ cannot give more information to the service provider. The results are the same if the overlapping grids are more than one. Therefore, the sensing area and the location of mobile user would not be exposed to the service provider and other entities.

\subsection{Data Confidentiality}
We aim to ensure that only the mobile users whose locations can match the sensing area have the capacity to recover the corresponding sensing task. In SPOON, the adversaries may be the service provider, unmatched mobile users and external attackers. To resist these adversaries, the task protection consists of two stages. In the first stage, the sensing task is encrypted by the customer under the public keys of the TA and the service provider; in the second one, the service provider partially decrypts the ciphertext using its secret key and then re-encrypts the result for the matched mobile users. Therefore, we demonstrate the task confidentiality in the following two procedures:
\begin{itemize}
  \item Firstly, the first-stage ciphertext should not be entirely decryptable for the service provider or the mobile users. To be specific, given the first-stage ciphertext ($c^*_1,c^*_2,c^*_3$) and two plaintexts $(task_1||u_1,task_2||u_2)$, if an adversary can distinguish which one out of $(task_1||u_1,task_2||u_2)$ is the plaintext of ($c^*_1,c^*_2,c^*_3$), we show how to construct a simulator $\mathcal{S}$ to solve the $q-$DBDHI problem \cite{Ateniese05}.

      Given the simplified $q-$DBDHI tuple $g,T_1=g^{z_1},T_2=g^{z_2} \in \mathbb{G}, Q \in \mathbb{G}_T$, the simulator $\mathcal{S}$'s goal is to determine whether $Q=\hat{e}(g,g)^{\frac{z_1}{z_2}}$ via interactions with the adversary. $\mathcal{S}$ sets $T=T_1$. The adversary possessing the secret key of the service provider, $\beta$, can query any chosen message $task||u$ to the simulator $\mathcal{S}$ to obtain the corresponding the ciphertext. Then, $\mathcal{S}$ picks two messages $(task_1||u_1,task_2||u_2)$ and a random bit $b\in \{0,1\}$ to compute the challenge $(c^*_1,c^*_2,c^*_3)=(S^{r_2}, T_2, (task_b||u_b)QH^{r_2})$, where $r_2$ is a random value chosen from $\mathbb{Z}_p$, and returns $(task_1||u_1,task_2||u_2)$ to the adversary, along with $(c^*_1,c^*_2,c^*_3)$. Finally, the adversary returns $\hat{b} \in \{0,1\}$ to $\mathcal{S}$. If $\hat{b}=b$, $\mathcal{S}$ can address the simplified $q-$DBDHI problem as  $Q\stackrel{?}{=}\frac{c^*_3}{(task_b||u_b)\hat{e}(c^*_1,h)^{-\frac{1}{\beta}}}$.

      The task confidentiality against the adversary, who possesses $\alpha$, also relies on the simplified $q-$DBDHI problem, given $h,T_1=h^{z_1},T_2=h^{z_2} \in \mathbb{G}, Q \in \mathbb{G}_T$, The proof is the same as that above with one difference that the challenge is $(c^*_1=T_2,c^*_2=T_1^{r_1},c^*_3=(task_b||u_b)QG^{r_1})$, where $r_1$ is a random value chosen from $\mathbb{Z}_p$. Finally, $\mathcal{S}$ can address the simplified $q-$DBDHI problem as $Q\stackrel{?}{=}\frac{c^*_3}{(task_b||u_b)\hat{e}(c^*_2,g)^{-\frac{1}{\alpha}}}$.

  \item Secondly, the sensing task should only be recovered by the matched mobile users from the second-stage ciphertext. To prevent unmatched mobile users from learning the content of sensing task, the service provider encrypts the sensing task with the temporary public key $\mu$ using the proxy re-encryption scheme \cite{Ateniese05}. Therefore, the security of the second-stage ciphertext can be reduced to the $q-$DBDHI assumption as well.
\end{itemize}

To guarantee the confidentiality of sensing reports, each mobile user employs the proxy re-encryption scheme \cite{Ateniese05} to encrypt $m_i$ under the temporary public key $u=g^k$, which is distributed to the mobile users along with the sensing task. The decryption key $k$ is kept by the customer secretly. Therefore, the confidentiality of $m_i$ directly depends on the sematic security of proxy re-encryption scheme, which can be reduced to the simplified $q-$DBDHI assumption \cite{Ateniese05}.

\subsection{Anonymity}
The anonymity of the mobile user is defined via the game in which the adversary cannot distinguish an honest mobile user out of two under the extreme condition that all other interactions are specified by the adversary. We prove that the mobile user's identity is preserved properly, unless the DDH assumption \cite{Cash08} does not hold. Specifically, if there exists an adversary $\mathcal{A}$ that can identify an honest mobile user out of two challenging identities, we show how to construct a simulator $\mathcal{S}$ to solve an instance of the DDH problem. That is, given a tuple $T_1,T_2,T_3,T_4 \in \mathbb{G}_T$, $\mathcal{S}$ can tell whether exists $(z_1,z_2)$, such that $T_2=T_1^{z_1}$, $T_3=T_1^{z_2}$, $T_4=T_1^{z_1z_2}$. $\mathcal{S}$ generates ($param, S, T$), picks two identities $(I_0,g^{a_0})$, $(I_1,g^{a_1})$, where $a_0,a_1 \in \mathbb{Z}_p$, and sends them to $\mathcal{A}$. $\mathcal{S}$ acts on behalf of the users $I_0$ and $I_1$ to register at the TA. $\mathcal{S}$ then interacts with $\mathcal{A}$ in the following interactions:
\begin{itemize}
  \item $\mathcal{S}$ acts as $I_0$ honestly to submit the location information. For $I_1$, in the $j$-th query, $\mathcal{S}$ randomly chooses $\mu_j \in\mathbb{G}$ and simulates the zero-knowledge proof $\mathcal{PK}_3$ to prove its identity interacting with $\mathcal{A}$.
  \item $\mathcal{S}$ honestly acts on behalf of $I_0$ to report the data. For $I_1$, $\mathcal{S}$ sets $H=T_1$, $\mathcal{G}=T_2$. For the $j$-th query, $\mathcal{S}$ randomly chooses $X_j$, $v_j \in \mathbb{Z}_p$ and computes $Y_j=T_1^{v_j}$, $Z_j=\hat{e}(g,g^{a_1})T_2^{X_jv_j}$. $\mathcal{S}$ simulates the zero-knowledge proof $\mathcal{SPK}$ and sends $(X_j,Y_j,Z_j,\mathcal{SPK})$ to $\mathcal{A}$, along with a random sensing report.
\end{itemize}
$\mathcal{S}$ picks a random bit $b \in \{0,1\}$. If $b=0$, $\mathcal{S}$ honestly reports the data acts as $I_0$. If $b=1$ and $\mathcal{S}$ randomly chooses $X_1\in \mathbb{Z}_p$ and calculates $\mathcal{G}=T_2$, $Y_1=T_3$, $Z_1=\hat{e}(g,g^{a_1})T_4^{X_1}$. Then, $\mathcal{S}$ simulates $\mathcal{SPK}$ and a sensing report, and sends them to $\mathcal{A}$. It is easy to see that the simulation is perfect if $log_{T_1}T_4=log_{T_1}T_2 log_{T_1}T_3$; otherwise, it contains no information about $I_0$ and $I_1$.

Finally, $\mathcal{A}$ returns $\hat{b}$. If $\hat{b}=b$, $\mathcal{S}$ can confirm that there exists $(z_1,z_2)$, such that $T_2=T_1^{z_1}$, $T_3=T_1^{z_2}$, $T_4=T_1^{z_1z_2}$. Thus, $\mathcal{S}$ resolves the DDH problem.

In the proof of customer's anonymity, a simulator $\mathcal{S}$ simulates the transcript of the zero-knowledge proof of the signature $(A,e,s)$, $\mathcal{PK}_2$, to interact with the adversary $\mathcal{A}$. Since $\mathcal{S}$ can perfectly simulates $\mathcal{PK}_2$, the adversary cannot obtain any identity information about the customer, such that it is impossible to distinguish an honest customer from two for $\mathcal{A}$. Therefore, the customer's anonymity can be fully guaranteed.

\subsection{Credit Balance}
Credit balance means that no one can own the credit points more than the initial credit points plus the credit points awarded by service providers. This is the most significant requirement for credit management from the respective of security. Assume $P_0$ be the initial credit points and $\theta_j$ be the earned points from the service provider in the $j$-th query. If the adversary $\mathcal{A}$ at most makes $\hat{R}$ reporting queries, and owns final credit points $P_f$, where $P_f>P_0+\sum_{j=1}^{\hat{R}} \theta_j$, while service providers do not identify the double-reporting, there must exist a simulator $\mathcal{S}$ to conduct a forgery attack on the underlying BBS+ signature \cite{Au06}.

Firstly, we assume that the zero-knowledge proofs $\mathcal{PK}_1$, $\mathcal{PK}_2$, $\mathcal{PK}_3$ and $\mathcal{SPK}$ are sound. That is, there exist extract algorithms $\mathcal{EX}_1$, $\mathcal{EX}_2$, $\mathcal{EX}_3$ and $\mathcal{EX}_S$ to obtain the witnesses of the zero-knowledge proofs, respectively.

Then, we show the simulator $\mathcal{S}$ that interacts with $\mathcal{A}$. $\mathcal{S} $ generates the public parameters $param$, the public keys $(T,S)$ and the secret keys $(\alpha, \beta)$, and is allowed to access the signature oracle $\mathcal{SO}$ to get the BBS+ signature of an input. $\mathcal{S} $ sends $(param, S, T)$ to $\mathcal{A}$ and interacts with $\mathcal{A} $ as follows:
\begin{itemize}
  \item $\mathcal{A}$ randomly chooses $C, C', \widehat{A} \in \mathbb{G}$, and generates the proof $\mathcal{PK}_1$ and sends them to $\mathcal{S}$. $\mathcal{S} $ extracts the witness $(s',t',a)$ from $\mathcal{PK}_1$ using $\mathcal{EX}_1$, and then picks a random credit point $P_0$ and queries the signature oracle $\mathcal{SO}$ to obtain $(A,e,s)$ and $(B,f,t)$. Finally, $\mathcal{S}$ calculates $s''=s-s'$, $t''=t-t'$ and $RK=\widehat{A}^{\frac{1}{\alpha}}$, and returns $(A,e,s'', B,f, t'', P_0, RK)$ to $\mathcal{A}$.
  \item For the $j$-th query, $\mathcal{A}$ picks a random $C'_j \in \mathbb{G}$ and executes $\mathcal{SPK}$ with $\mathcal{S}$. $\mathcal{S}$ utilizes $\mathcal{EX}_S$ to extract the witness $(B_j,f_j,t_j,t'_j,a_j,I_j,P_j,v_j)$. If $(B_j,f_j,t_j)$ is not an output of $\mathcal{SO}$, it is a forgery of the BBS+ signature. Otherwise, $\mathcal{S}$ queries $\mathcal{SO}$ to obtain a signature $(B_j,f_j,t_j)$ on input $(a_j,P_j+\theta_j,I_j)$. $\mathcal{S}$ receives $(B_j,f_j,t_j)$ and computes $t''_j=t_j-t'_j$, and returns $(B_j,f_j,t''_j)$ to $\mathcal{A}$.
\end{itemize}

Finally, assume $\mathcal{A}$ executes $\hat{R}$ queries. $\mathcal{A}$ wins the game if it can prove $P_f>P_0+\sum_{j=1}^{\hat{R}} \theta_j$. However, if $P_f>P_0+\sum_{j=1}^{\hat{R}}  \theta_j$, $\mathcal{A}$ must have conducted a forged BBS+ signautre or double-reported the data. While the BBS+ signature is secure under the $q-$SDH assumption \cite{Au06}, $\mathcal{A}$ cannot forge a BBS+ signature, unless the $q-$SDH assumption \cite{Au06} does not hold. If $\mathcal{A}$ double-reports the sensing data, it must generate another $\widetilde{Z}_i$, which is unequal to the previous $Z_i$, in the same time slot. Due to the soundness of zero-knowledge proof protocol, $Z_i=\hat{e}(g,\widehat{A}_i)\mathcal{G}^{X_iv_i}$ is the only valid $Z_i$ to accompany the specific report identified by $X_i$ and $Y_i$. Since $X_i$ should be different for two reports, $\hat{e}(g,\widehat{A}_i)$ would be obtained as long as the proof is valid. We assume the proof $\mathcal{SPK}$ is sound. Thus, the success probability of double-reporting for $\mathcal{A}$ is negligible. Therefore, the probability to obtain $P_f>P_0+\sum_{j=1}^{\hat{R}} \theta_j$ is negligible if the $q-$SDH assumption holds.

\subsection{Greedy User Tracing}
Greedy user tracing consists of two objectives, namely, slandering and hiding.  Slandering means that an attacker cannot slander an honest mobile user, and hiding means that a greedy user must be identified by the TA. For the slandering, the attacker releases pieces of reporting transcripts that can link to other reports submitted by an honest mobile user. It is infeasible for the attacker to compute the tracing information about an honest mobile user since the proof $\mathcal{SPK}$ is sound. Therefore, no attacker enables to slander an honest mobile user. In terms of the hiding, the attacker is required to generate different pieces of tracing information without being traced. However, it is impossible for a greedy mobile user to compute $Z_i$ if the pseudo-random function $\mathcal{F}$ is correct.

\section{Extension} \label{sec5}
In this section, we propose an approach to evaluate the trust levels of sensing reports and a new location matching mechanism to achieve communication-efficient task allocation for mobile crowdsensing.

\subsection{Evaluation on Trust Level}
The service provider uploads $w$ sensing reports to the customer, and the customer evaluates the trust level of each report, and distributes credit points to mobile users. Sensing reports have distinct trustworthiness due to the various intelligence of data sources. Furthermore, some mobile users may forge the sensing data or deliver ambiguous, biased data to gain credit points by cheating. Therefore, we propose a fair trust evaluation mechanism as follows:

\begin{itemize}
  \item The customer generates the weights of sensing data associated with grids in the sensing region $\omega_{z\{z \in \mathcal{L}\}} \in (0,1]$, such that $\sum_{\{z \in \mathcal{L}\}} \omega_z=1$, and divides $w$ sensing reports into $|\mathcal{L}|$ groups, where $|\mathcal{L}|$ means the number of the grids in the sensing area. If the data in a sensing report are collected from several grids, this report is in the groups associated with these grids meanwhile.

  \item For each sensing report in a group $z \in \mathcal{L}$, the customer computes the similarity $V_{i,z}$. If a sensing report is significantly different from the others in the same group (e.g., an opposite result), its similarity is set to be a negative value $V_{i,z} \in [-\gamma,0]$. Otherwise, the customer sets a positive similarity $V_{i,z} \in (0,\gamma]$ for the report.
  \item For each sensing report in a group $z \in \mathcal{L}$, the customer computes $\rho_{i,z}=V_{i,z} Q_i$ and $Exp_z=\sum_{i\in \mathcal{Z}} \rho_{i,z}$, where $\mathcal{Z}$ denotes the set of the sensing reports in the group $z$. Then, the customer sets the trust level of the sensing report to be $\epsilon_{i,z}=(\frac{\rho_{i,z}}{Exp_z})\omega_z \gamma$.
 \item If the report only contains the data collected from one grid, its trust level is $\epsilon_i=\epsilon_{i,z}$; otherwise, the trust level is set to be the average of the trust levels for all the grids where $m_i$ is collected as $\epsilon_i=AVE_z(\epsilon_{i,z})$.
\end{itemize}

\subsection{Efficiency-enhanced Task Allocation}
In SPOON, we use the matrices $\widehat{L}_{m\times n}$ and $\widetilde{L}_{m\times n}$ to represent the sensing area of the task and the location of the mobile user, respectively. To prevent attackers from acquiring the location information, $\widehat{M}_{m\times n}$, $\widetilde{M}_{m\times n}$ are exploited to randomize $\widehat{L}_{m\times n}$, $\widetilde{L}_{m\times n}$. Although this approach is computationally efficient and the service provider cannot identify the exact region, the service provider has to define its service region in service setup phase and the communication overhead in task allocation phase is a little heavy, since both the customer and the  mobile user are required to transmit the matrices $\widehat{N}_{n\times n}$, $\widetilde{N}_{n\times n}$ to the service provider. To reduce the communication cost, we propose a communication-efficient location matching mechanism by employing the BGN encryption \cite{Boneh05}.

\begin{figure}
\centerline{\includegraphics[width=0.25\textwidth]{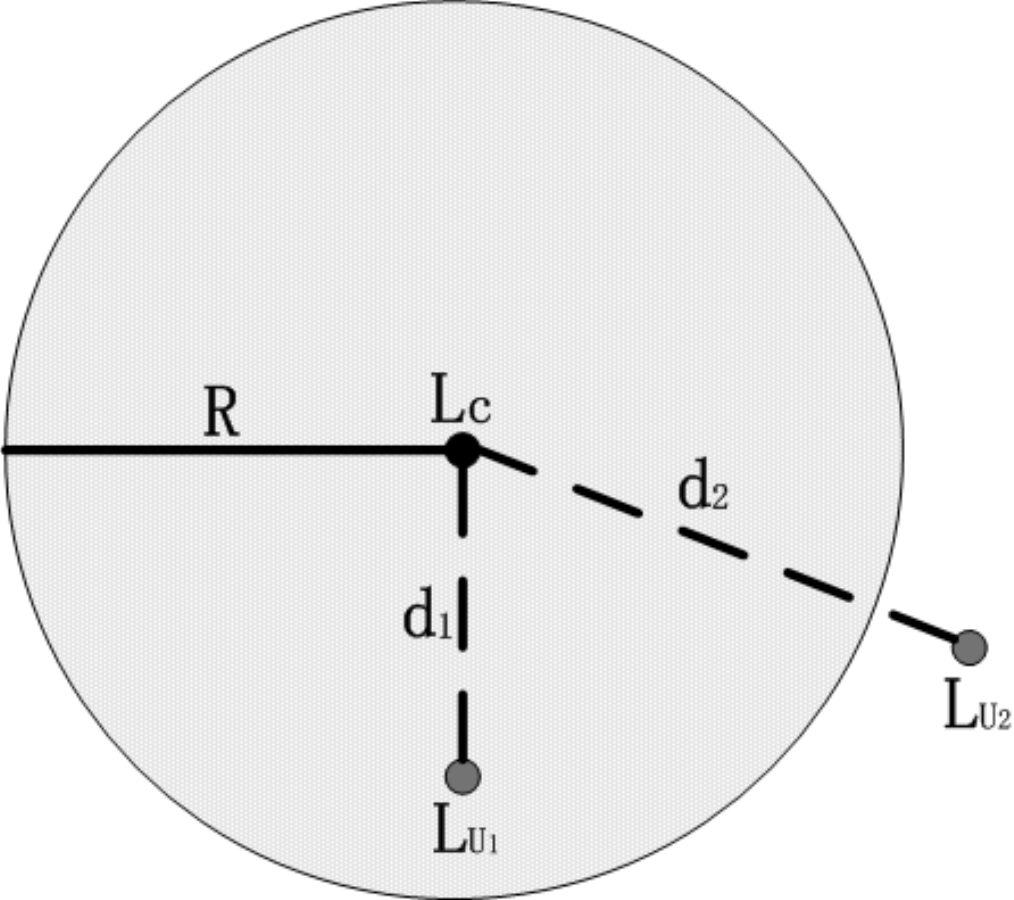}}
\caption{Sensing Circle and Location.}
\label{fig:four}
\end{figure}

In the service setup phase, apart from generating $param$ and the secret-public key pair $(\alpha, T)$, the TA setups the BGN encryption. It chooses two random $\lambda$-bit primes $q_1,q_2$, sets $n=q_1q_2$, and generates two bilinear groups $\mathbb{G}_1,\mathbb{G}_2$ of order $n$ that satisfies the bilinear map $\tilde{e}: \mathbb{G}_1 \times \mathbb{G}_1\rightarrow \mathbb{G}_2$. It also picks a random generator $l \in \mathbb{G}_1$ to compute $l_1=l^{q_2}$. Thus, the public key of the TA is $(n, \mathbb{G}_1, \mathbb{G}_2, l, l_1, T)$ and the secret key is $(\alpha, q_1)$.

When a customer with  $(A,e,s,B,f,t,a,I,P,\widehat{A}, RK)$ has a sensing task $ST=(task, expires, area,\gamma,w)$ to be allocated to the mobile users, it generates $(c_1,c_2,c_3,u,\mathcal{PK}_2)$, following the steps given in IV. C. The sensing area $area$ is defined as a circle, which is uniquely identified by a center $L_c=(L_{cx},L_{cy})$ and a radius $R$, where $L_{cx}$ is the longitude and $L_{cy}$ is the latitude. As shown in Fig. 4, assume the geographical location of a mobile user $U_1$ is $L_{u_1}$. If the distance between $L_{u_1}$ and $L_c$ is shorter than $R$, $U_1$ is located in the sensing area; otherwise, it is out of the crowdsensing region. To protect the sensing region, the customer picks three random values $r'_x, r'_y, r'_r \in \mathbb{Z}_n$ and computes $C_x=l^{L_{cx}}l_1^{r'_x}$, $C_y=l^{L_{cy}}l_1^{r'_y}$, and $C_R=l^{R}l_1^{r'_r}$. The customer sends $(c_1,c_2,c_3,expires,\mathcal{PK}_2, C_x,C_y,C_R,\gamma,w)$ to the service provider and the latter releases the task $ST$ as described in IV. C.

If a mobile user $U_i$ wants to perform sensing tasks, it chooses a random $\nu \in \mathbb{Z}_p$ to compute $\mu=h^{\nu}$. Then, $U_i$ retrieves the location information $L_{u_i}=({L_{ix},L_{iy}})$ from the GPS device or the access point, and encrypts it using the public key of the TA as follows: pick random values $r''_x, r''_y \in \mathbb{Z}_n$ to calculate $U_x=l^{L_{ix}}l_1^{r''_x}$, $U_y=l^{L_{iy}}l_1^{r''_y}$. Finally, $U_i$ generates $\mathcal{PK}_3$ and sends $(\mu, U_x,U_y, \mathcal{PK}_3)$ to the service provider.

Upon receiving $(\mu, U_x,U_y, \mathcal{PK}_3)$, the service provider firstly determines whether a sensing task's sensing region covers the location of this user. For each unexpired task, the service provider computes $X'=\tilde{e}(\frac{C_x}{U_x},\frac{C_x}{U_x})$, $Y'=\tilde{e}(\frac{C_y}{U_y},\frac{C_y}{U_y})$ and $Z'=X'Y'$, and sends $Z'$ to the TA, along with $(C_R, num)$ for every task. The TA decrypts $Z'$ and $C_R$ to recover $d_i$ and $R$, respectively, and checks whether $d_i< R$ to find the set of matching sensing tasks, and returns the task numbers $num$ to the service provider. Then, the service provider generates $c_4$ and releases $(num,c_2, c_3,c_4, expires, \gamma)$ for $U_i$. Finally, $U_i$ obtains the sensing task and collects data.

This location matching mechanism is built from the BGN encryption and the homomorphic property is utilized to compute the distance from the circle center to the location of mobile users. The security of this mechanism can be reduced to the sematic security of the BGN encryption scheme. Moreover, the customer is required to send $(C_x,C_y,C_R)$ and the mobile user is needed to deliver $(U_x,U_y)$ to the operation center, which are shorter than $\widehat{N}_{n\times n}$ and $\widetilde{N}_{n\times n}$

\begin{table*}
\caption{Computational Overhead of SPOON}
\centering
\begin{tabular}{|c|c|c|c|c|c|c|c|c|c|c|}
\hline
\multirow{2}{*}{Phase} &\multicolumn{2}{|c|}{User Registration} & \multicolumn{3}{|c|}{Task Allocation} & \multicolumn{3}{|c|}{Data Reporting}  & \multicolumn{2}{|c|}{Credit Assignment} \\
\cline{2-11}
& Authority& User & Customer& Provider& User& Customer&Provider & User & Provider & User \\
\hline\hline
Point Multiplication & 16 & 19 & 11 & 12 & 9 &  0 & 19 & 25 & 3&5  \\
\hline
Point Addition & 12 & 13 & 5 & 8 & 5  & 0 & 14 & 16 & 3&5  \\
\hline
Bilinear Map & 0 & 4 & 1 & 1& 2  & 1 & 5 & 2 & 0&2  \\
\hline
Exponentiation in $\mathbb{G}_T$   & 0 & 0 & 6 & 15& 8 & 1 & 19 &15 & 0&0  \\
\hline
Running Time (ms)  & 83.429 & 293.372 & 100.123 & 138.529& 154.980  & 56.415 & 197.324 &203.129 & 15.643&130.448  \\
\hline
\end{tabular}
\end{table*}

\section{Performance Evaluation}\label{sec6}
In this section, we evaluate the performance of our SPOON in terms of computational and communication overheads, and analyze privacy rate and accuracy rate for credit management.

\subsection{Computational Overhead}
We demonstrate the computational overhead of our SPOON by counting the number of the time-consuming  cryptographic operations, such as point multiplication, point addition, bilinear map and exponentiation in $\mathbb{G}_T$. Here we only show four kinds of operations because other operations, e.g., multiplication in $\mathbb{G}_T$, addition, multiplication and inverse operations in $\mathbb{Z}_p$, are not comparable with these four operations. Besides, since the bilinear map is the most time-consuming operation in cryptographic calculations, we utilize the pre-processing technique to reduce the computational burden for each entity. Specifically, the TA pre-computes the bilinear maps $\{E_i\}_{i=0}^4,\{F_i\}_{i=0}^4,K, K'_0,\{K_i\}_{i=0}^3$ in service setup phase as shown in Appendix A, and the bilinear maps $\{\hat{e}(g,\widehat{A}_i)\}_{i=0}^N$ in user registration phase, where $N$ is the number of registrants. The mobile user $U_i$ also can pre-compute $\hat{e}(g,\widehat{A}_i)$ in user registration phase. Table III shows the number of the operations executed by each entity in each phase of SPOON, respectively.

We also conduct an experiment to show the efficiency of SPOON. The operations of TA and service provider are performed on a notebook with Intel Core i5-4200U CPU, the clock rate is 2.29GHz and
the memory is 4.00 GB. The operations of customers and mobile users are run on HUAWEI MT2-L01 smartphone with Kirin 910 CPU and 1250M memory. The operation system is Android 4.2.2 and the toolset is Android NDK r8d. We use MIRACL library 5.6.1 to implement number-theoretic based methods of cryptography. The Weil pairing is utilized to realize the bilinear pairing. The parameter $p$ is approximately 160 bits and the elliptic curve is defined as $y=x^3+1$ over $\mathbb{F}_q$, where $q$ is 512 bits. The execution time of each entity in every phase of SPOON is shown in Table III. The running time is less than 300 ms for each entity. Therefore, our SPOON is quite efficient to be deployed on mobile devices.

\subsection{Communication Overhead}
We show the communication burden of all entities in SPOON. The public parameters are set the same as those in the experiment, that is, $|p|$=160 bits and $|q|$=512 bits. In user registration phase, a registrant, either customer or mobile user, sends a registering request $(I,C,C',\widehat{A},\mathcal{PK}_1)$ to the TA, which is $|I|+2176$ bits, where $|I|$ is the binary length of the identity, and the TA returns $(A,B, s'',t'', e, f,P_0, RK)$ to the registrant, whose binary length is $|P_0|+2176$ bits, where $|P_0|$ is the binary length of credit point. In task allocation, the customer uploads $(c_1,c_2,c_3,expires,\widehat{N}_{n \times n}, \gamma, w, \mathcal{PK}_2)$ and the mobile user sends $(\mu, \widetilde{N}_{n \times n},\mathcal{PK}_3)$ to the service provider, which are $4512+160n^2+|expires|+|\gamma|+|w|$ bits and $2976+160n^2$ bits, respectively. The service provider responds $(num,c_2,c_3,c_4, expires,\gamma)$, which is $2560+|num|+|expires|+|\gamma|$ bits, to a matched mobile user or false, 1 bit, to an unmatched one. After the mobile user obtains the sensing data, it generates the sensing report $(num, D_i, D'_i, C'_i,X_i,Y_i,Z_i,Q_i,\tau_j,\mathcal{SPK})$ to the service provider, which is $8864+|num|+|P_0|+|\tau_j|$ bits. The service provider needs to send 1024-bit $W$ to the TA if a mobile user double-submits data, and then sends $w$ sensing reports $(num, D_i, D'_i,Y_i, Q_i,\tau_j)$, which is of binary length $w*(2560+|num|+|P_0|+|\tau_j|)$ to the customer. Finally, the customer returns $(1024+|\gamma|)$-bit $(\epsilon_i,Y_i)$ to the service provider for each report and the service provider sends $(B_i,t''_i,f_i,\theta_i,Y_i)$ to every mobile user, which is $1856+|P_0|$ binary bits.

\subsection{Credit Analysis}
To prevent credit points from disclosing to other entities, each mobile user claims a threshold $Q_i$, which is less than its exact credit point $P_i$, such that the cloud provider can select the sensing reports based on the claimed thresholds. In this way, neither the cloud provider nor the customer enables to learn the precise credit points of mobile users. Unfortunately, this method reduces the accuracy of report selection as the cloud provider may select a sensing report of the mobile user, whose threshold is larger than others', while the credit point has the opposite result. On the other hand, customers may prefer mobile users to choose the thresholds that are approximate to their credit points, while the privacy of mobile users are sacrificed. Therefore, it seems to be impossible to reconcile the contradiction between privacy and accuracy, because they have an opposite of trends.

To balance this trade-off, it is critical to find a reasonable strategy for mobile users to determine the thresholds. We define four parameters to evaluate privacy and accuracy in credit claiming. Specifically, accuracy rate A denotes the maximum probability of a given threshold in the selected reports can possess top-$w$ credit points in sensing reports. Accuracy rate B denotes the maximum probability that a given credit point in the sensing reports is larger than the minimum threshold in the selected reports. Privacy rate A means the probability that a given sensing report, whose credit point is larger than the minimum of thresholds in selected reports, has top-$w$ credit point in all sensing reports. Privacy rate B means the probability that a given sensing report would be selected by the service provider, whose credit point is larger than the minimum of thresholds in selected reports. To determine how the threshold choosing strategy impacts the defined privacy and accuracy rates, we simulate the credit points of mobile users on Matlab and use different threshold choosing strategies to compute the accuracy rates and the privacy rates. The simulation results are illustrated in Fig. 5 and Fig. 6. We set the number of the mobile users to be 1000 in Fig. 5 and the number of the selected reports to be 100 in Fig. 6. We compare three threshold choosing strategies, the first one is basing on uniform distribution; the second one is basing on Gaussian distribution, in which the mean is three quarters of credit points and the standard deviation is one quarter; the last one is basing on Gaussian distribution, where the mean and the standard deviation are one quarter of credit points. The second strategy can achieve the highest accuracy and the third strategy can achieve the best privacy preservation on credit points in three strategies.

\begin{figure}
\centering
\subfigure[Accuracy Rate A]{
\label{Fig61}
\includegraphics[width=0.23\textwidth]{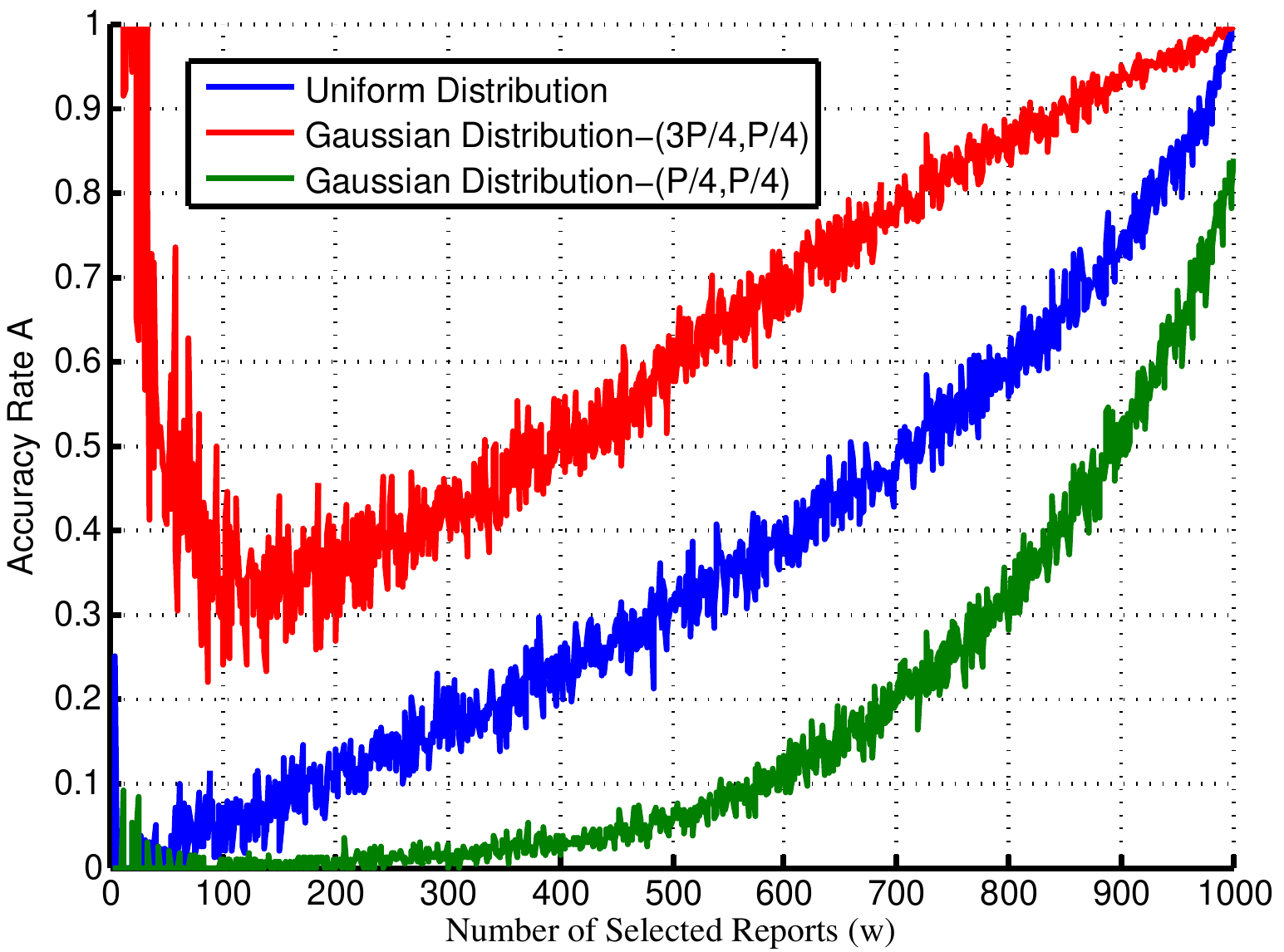}}
\subfigure[Accuracy Rate B]{
\label{Fig62}
\includegraphics[width=0.23\textwidth]{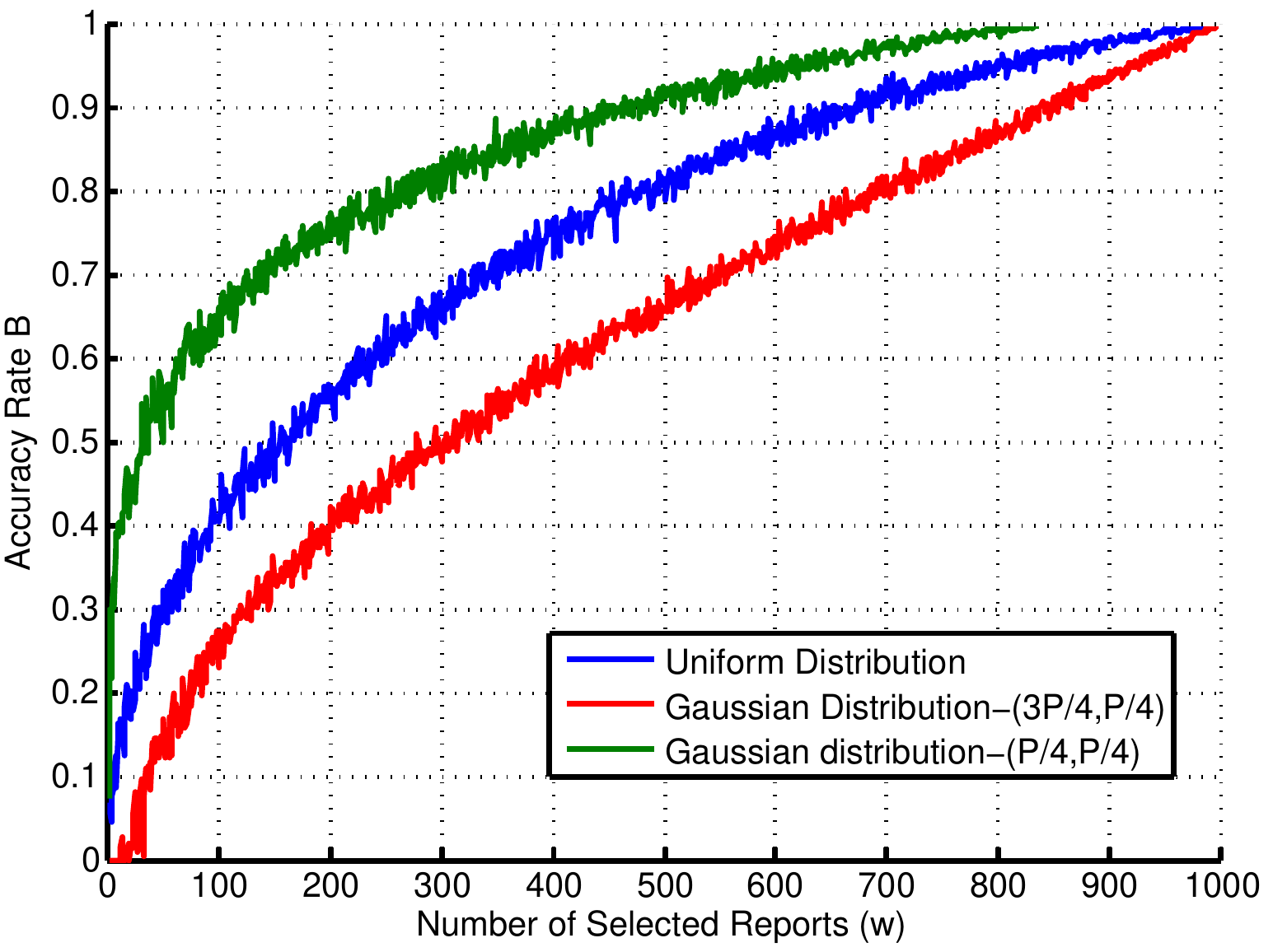}}
\subfigure[Privacy Rate A]{
\label{Fig63}
\includegraphics[width=0.23\textwidth]{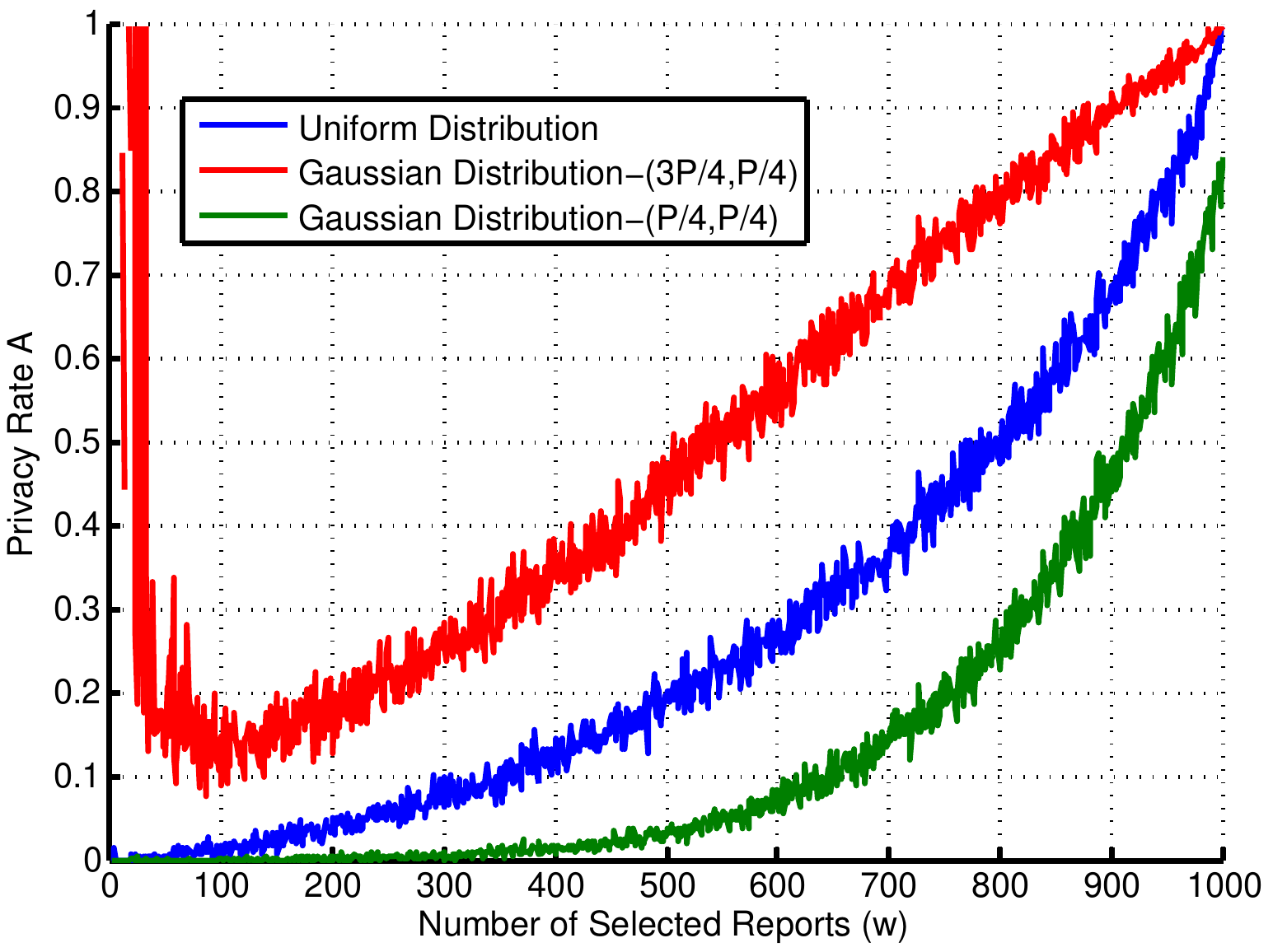}}
\subfigure[Privacy Rate B]{
\label{Fig64}
\includegraphics[width=0.23\textwidth]{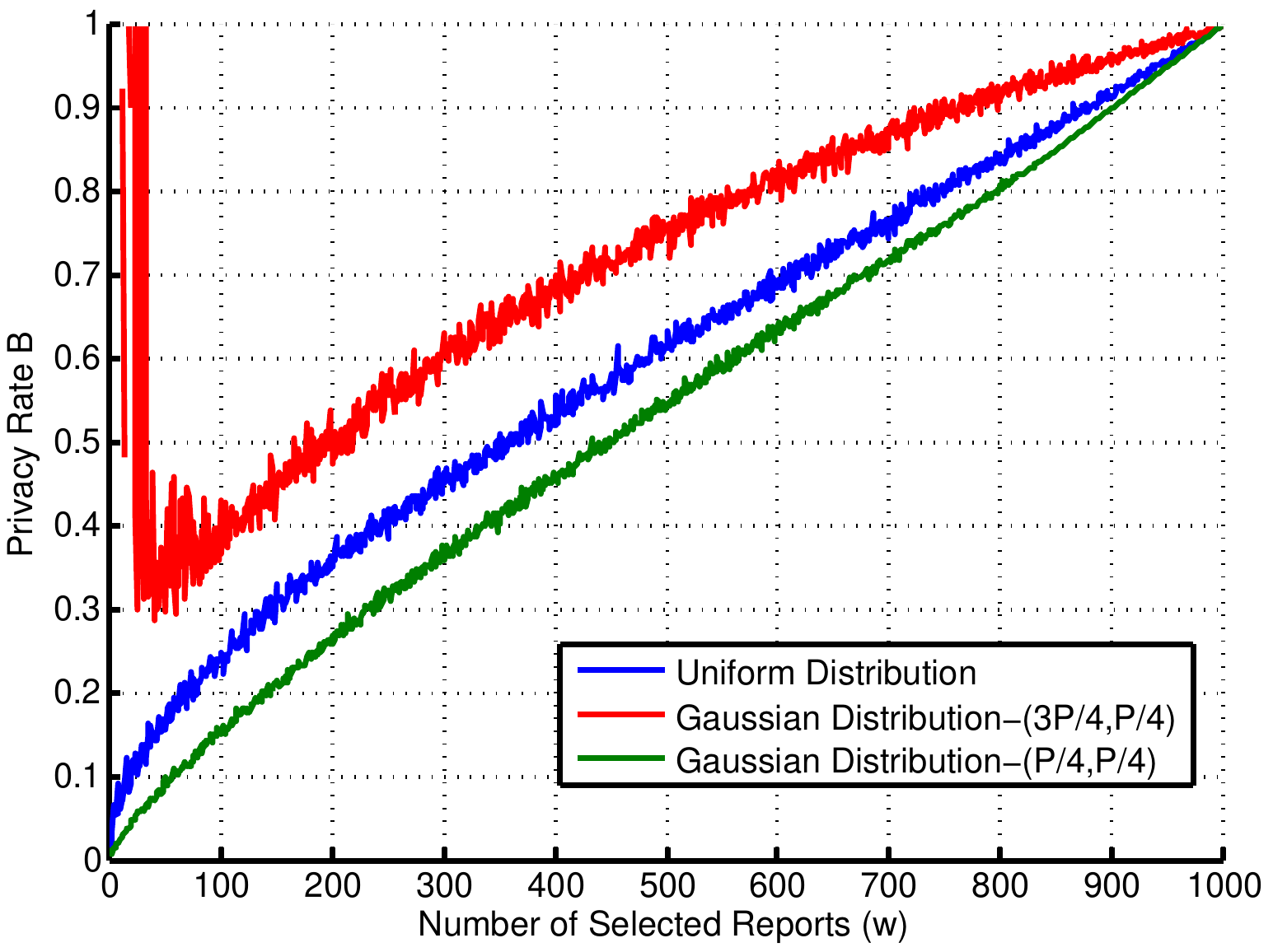}}
\caption{Accuracy and Privacy Rates with $N$=1000}
\label{Fig6}
\end{figure}

\section{Conclusions} \label{sec8}
In this paper, we have proposed a strong privacy-preserving mobile crowdsensing scheme with credit management to balance the trade-off between privacy preservation and task allocation. The service provider is allowed to select mobile users to perform sensing tasks according to the sensing areas of tasks and the geographic locations of mobile users, and select the sensing reports based on the credit points of mobile users. The sensitive information, including identities, locations, credit points, sensing tasks and sensing reports are preserved for mobile users and customers during task allocation and report selection. Furthermore, no trusted third party is required to achieve the credit management for mobile users.
In the future work, we will design a privacy-preserving context-aware task allocation framework for mobile crowdsensing.

\appendix{}
\begin{center}
Details of $\mathcal{PK}_1 \sim \mathcal{PK}_3$ and $\mathcal{SPK}$
\end{center}

$\mathcal{SPK}$ demonstrates that the number of credit points $P_i$ that a mobile user has is larger than the claimed credit threshold $Q_i$. Thanks to the zero-knowledge range proof due to Camenisch et al. \cite{Camenisch08}, the mobile user can prove that the value $P_i-Q_i$ is non-negative. We fix the internal of $[0,V]$, where $V$ is chosen by the service provider and is large enough compared with the credit points of all mobile users. Utilizing the efficient interval proofs in \cite{Allen14}, the mobile user demonstrates that $P_i-Q_i$ is one element in the interval $[0,V]$. To instantiate the zero-knowledge proofs $\mathcal{PK}_1 \sim \mathcal{PK}_3$ and $\mathcal{SPK}$, in the setup phase, the service provider generates some auxiliary parameters $y,y_1,y_2 \in \mathbb{G}$, $\eta =y^{\varphi}$ for a randomly chosen value $\varphi \in \mathbb{Z}_p$, and computes $\phi_{\iota}=y^{\frac{1}{\iota+\varphi}}$, for each $\iota=0$ to $V$. To improve the efficiency, the TA can pre-compute $E_0=\hat{e}(g_0,g)$, $E_1=\hat{e}(g_1,g)$, $E_2=\hat{e}(g_2,g)$, $E_3=\hat{e}(g_3,g)$, $E_4=\hat{e}(g_2,S)$, $F_0=\hat{e}(h_0,h)$, $F_1=\hat{e}(h_1,h)$, $F_2=\hat{e}(h_2,h)$, $F_3=\hat{e}(h_3,h)$, $F_4=\hat{e}(h_4,h)$, and the service provider pre-computes $K=\hat{e}(y,y)$, $K_0=\hat{e}(y_1,T)$, $K'_0=\hat{e}(y_1,S)$ $K_1=\hat{e}(y_1,h)$, $K_2=\hat{e}(y_1,y)$, $K_3=\hat{e}(y_1,\eta)$. These parameters are included in the public parameters as $param \cap \{E_0,E_1,E_2,E_3,E_4,F_0,F_1,F_2,F_3,F_4\}$. The service provider also releases the parameters $\{S, L_{m \times n}, y,y_1,y_2,\eta,\{\phi_{\iota}\}_{\iota \in [0,V]}, K,K_0,K'_0,K_1,K_2,K_3\}$. To deduce the number of interactions in zero-knowledge proofs, we utilize the Fiat-Shamir transformation, where the hash function $\mathcal{H}$ can be viewed as a random oracle. The details of $\mathcal{PK}_1 \sim \mathcal{PK}_3$ and $\mathcal{SPK}$ are shown below.

\begin{figure}
\centering
\subfigure[Accuracy Rate A]{
\label{Fig61}
\includegraphics[width=0.23\textwidth]{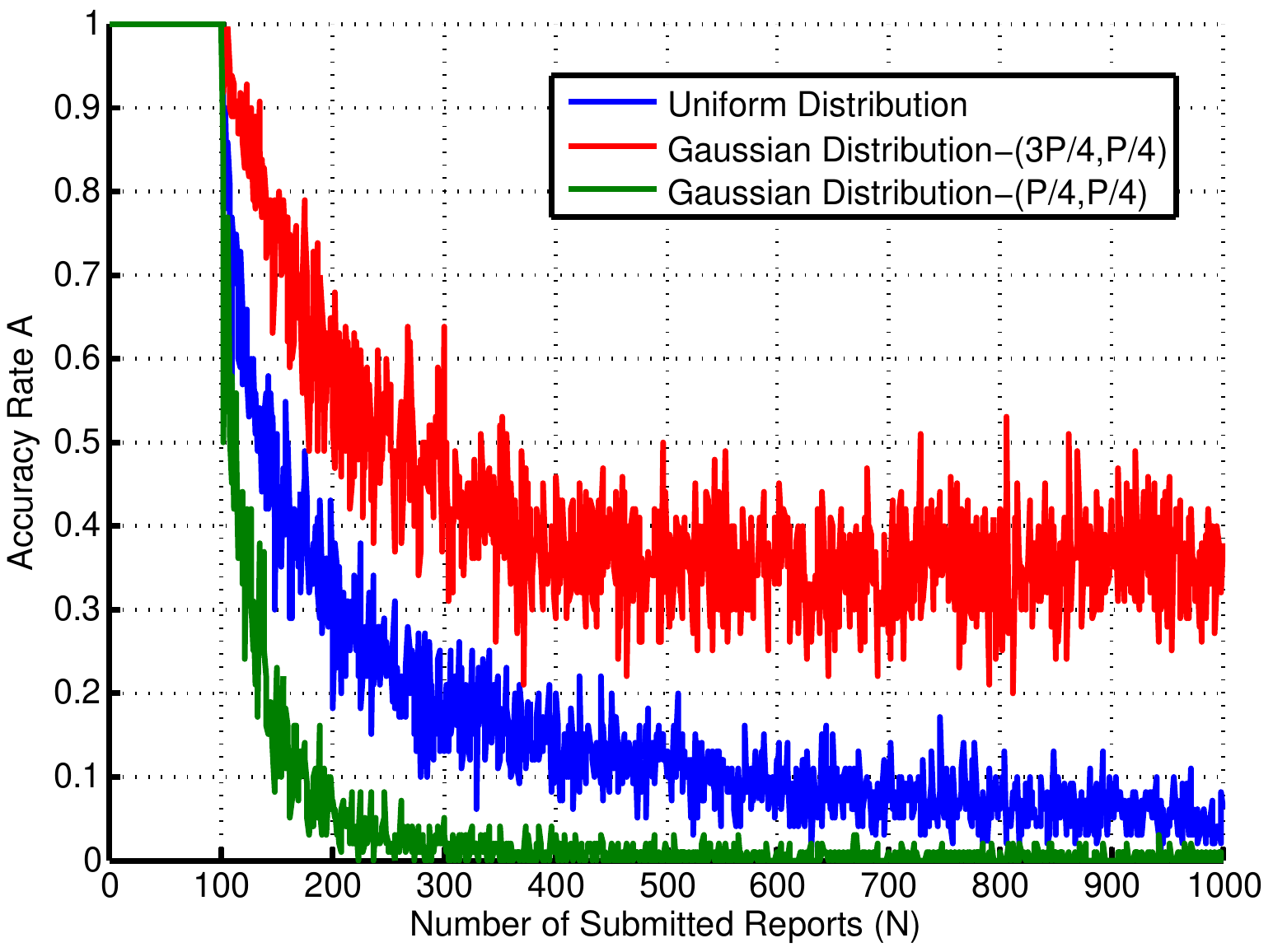}}
\subfigure[Accuracy Rate B]{
\label{Fig62}
\includegraphics[width=0.23\textwidth]{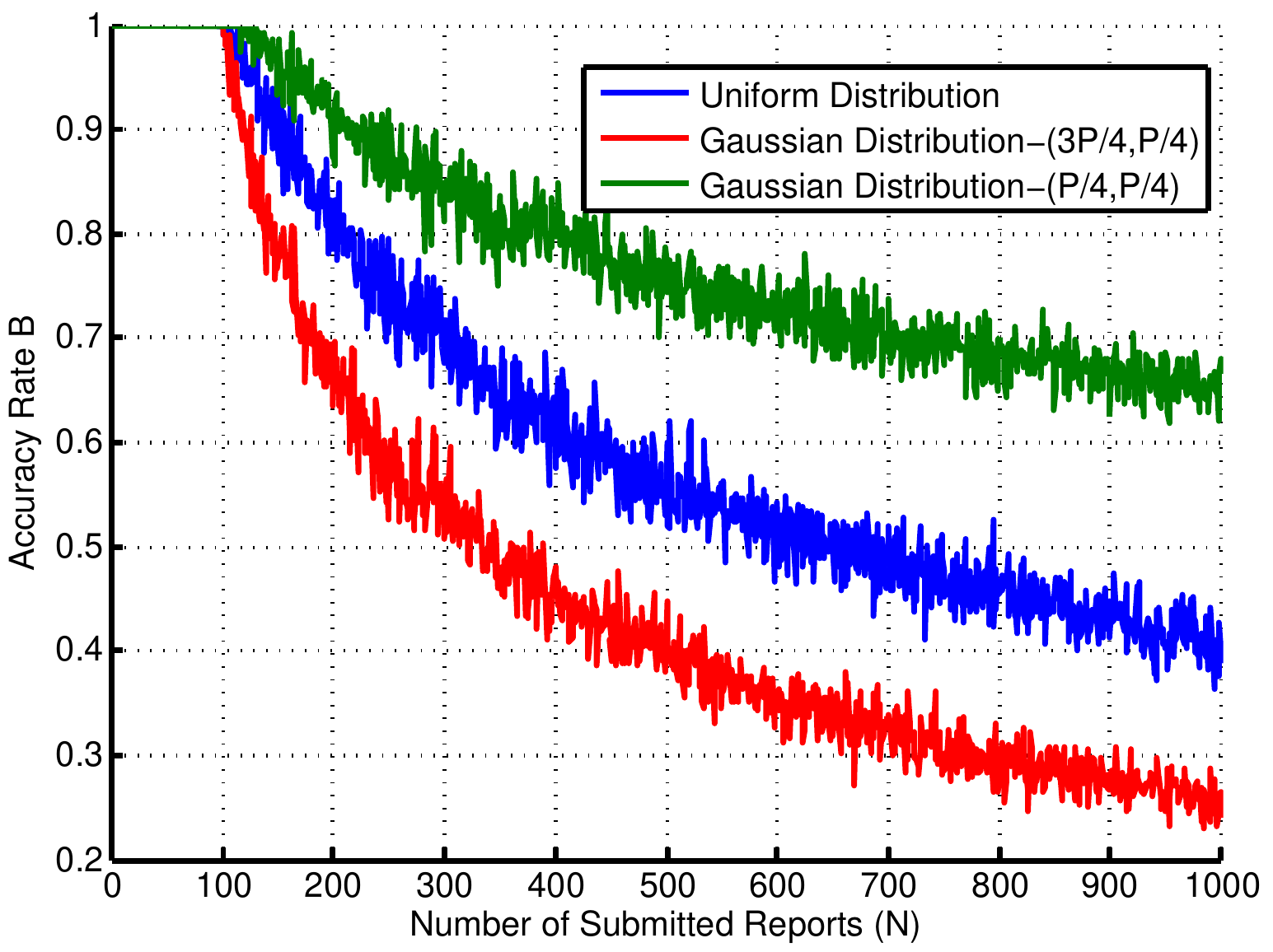}}
\subfigure[Privacy Rate A]{
\label{Fig63}
\includegraphics[width=0.23\textwidth]{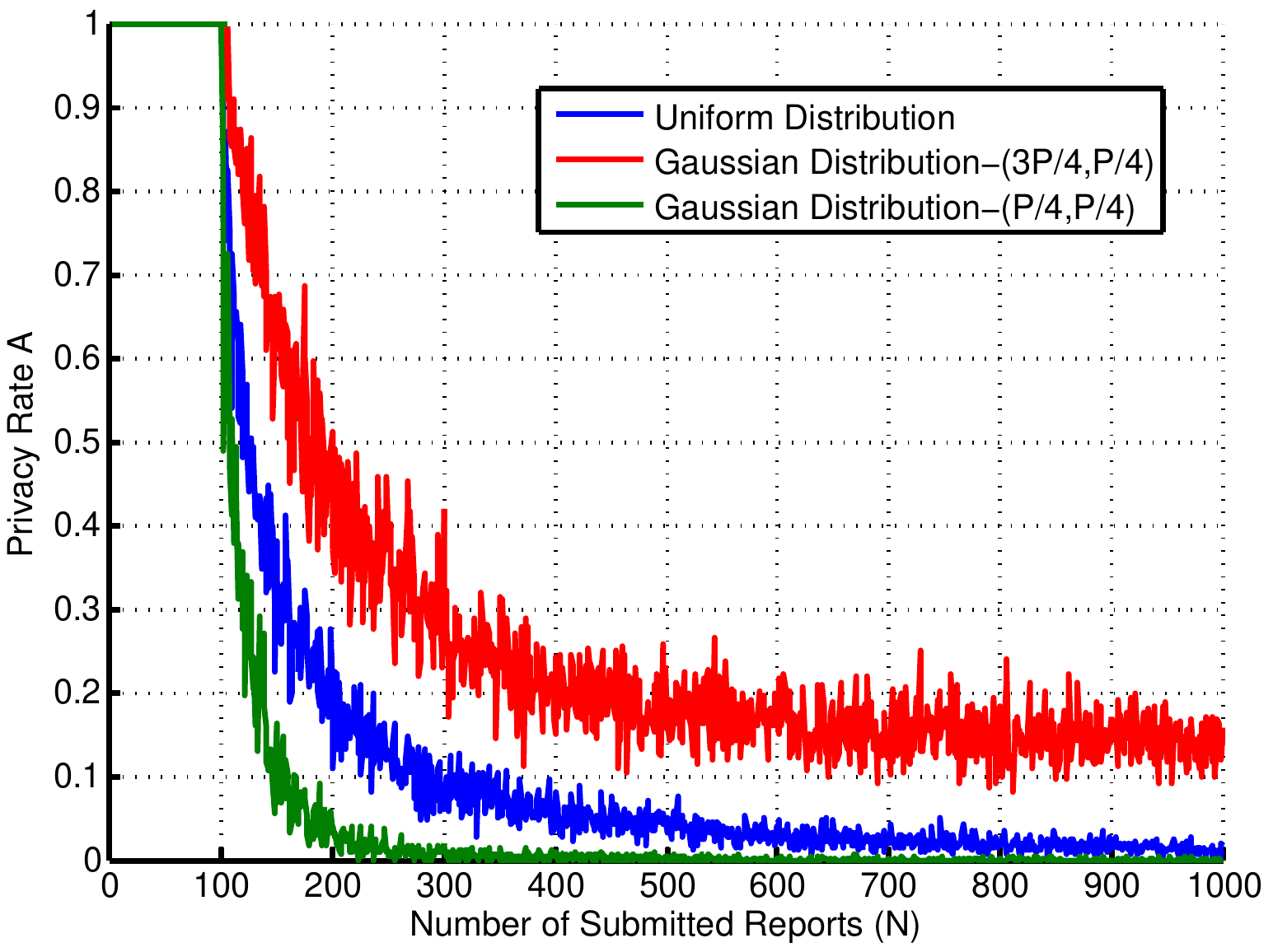}}
\subfigure[Privacy Rate B]{
\label{Fig64}
\includegraphics[width=0.23\textwidth]{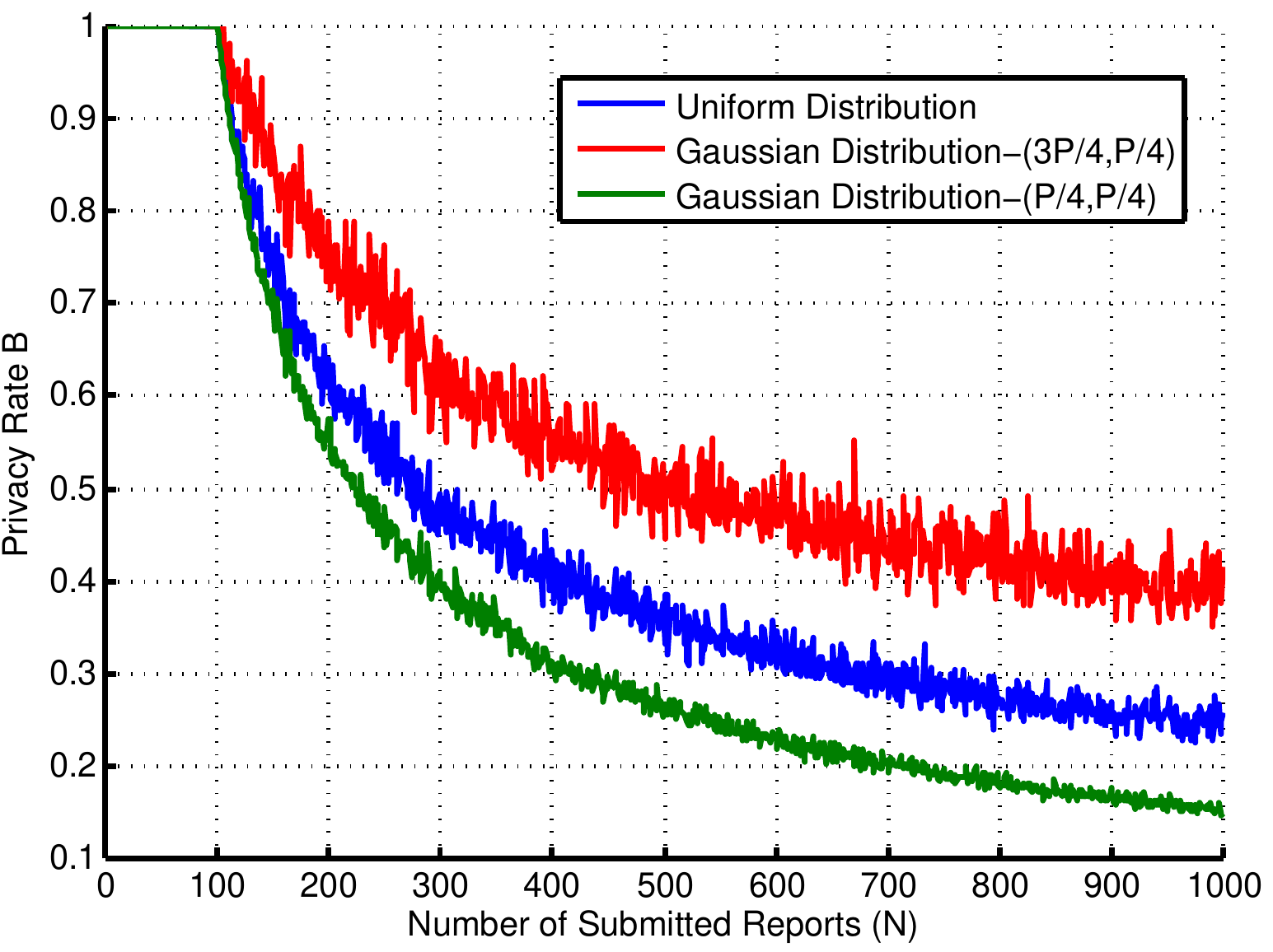}}
\caption{Accuracy and Privacy Rates with $w$=100}
\label{Fig6}
\end{figure}

\noindent $\mathcal{PK}_1\{(s',t',a): C=g_1^{s'}g_2^a \land C'=h_1^{t'}h_2^a\land \widehat{A}=g^a\}.$
\begin{enumerate}
  \item The TA sends a random challenge $R\in \mathbb{Z}_p$.
  \item The registrant randomly chooses $\rho_{s'},\rho_{t'},\rho_{a} \in \mathbb{Z}_p$ and computes $T_1=g_1^{\rho_{s'}}g_2^{\rho_{a}}$, $T_2=h_1^{\rho_{t'}}h_2^{\rho_{a} }$ and $T_3=g^{\rho_{a}}$.
  \item The registrant computes $c=\mathcal{H}(T_1,T_2,T_3,R)$.
  \item The registrant computes $z_{s'}=\rho_{s'}-cs'$, $z_{t'}=\rho_{t'}-ct'$, $z_{a}=\rho_{a}-ca$, and sends $c, z_{s'}, z_{t'}, z_{a}$ to the TA.
  \item The TA computes $T'_1=C^c g_1^{z_{s'}}g_2^{z_{a}}$, $T'_2={C'}^c h_1^{z_{t'}}h_2^{z_{a} }$, $T'_3=\widehat{A}^cg^{z_{a}}$ and accepts the proof if $c=\mathcal{H}(T'_1,T'_2,T'_3,R)$; otherwise, rejects.
\end{enumerate}

\noindent $\mathcal{PK}_2\{(A,e,s,a, I): \hat{e}(A,Tg^e){=}\hat{e}(g_0g_1^sg_2^ag_3^I,g)\}.$
\begin{enumerate}
  \item The service provider sends a random challenge $R\in \mathbb{Z}_p$.
  \item The customer randomly chooses $r_1,r_2 \in \mathbb{Z}_p$ to compute $B_1=g_1^{r_1}g_2^{r_2}$, $B_2=B_1g_2^{r_1}$, $\delta_1=r_1e$, $\delta_2=r_2e$. The customer picks random $\rho_{r_1},\rho_{r_2},\rho_{\delta_1},\rho_{\delta_2}, \rho_{e},\rho_{s},\rho_{a}, \rho_{I} \in \mathbb{Z}_p$ and computes $T_1=g_1^{\rho_{r_1}}g_2^{\rho_{r_2}}$, $T_2=B_1^{-\rho_{e}}g_1^{\rho_{\delta_1}}g_2^{\rho_{\delta_2}}$, $T_3=\hat{e}(B_2,g)^{-\rho_{e}}E_1^{\rho_{s}}E_2^{\rho_{a}}E_3^{\rho_{I}}E_4^{\rho_{r_1}}E_2^{\rho_{\delta_1}}$.
  \item The customer computes $c=\mathcal{H}(T_1,T_2,T_3,R)$.
  \item The customer computes $z_{r_1}=\rho_{r_1}-cr_1$, $z_{r_2}=\rho_{r_2}-cr_2$, $z_{\delta_1}=\rho_{\delta_1}-c\delta_1$, $z_{\delta_2}=\rho_{\delta_2}-c{\delta_2}$, $z_{e}=\rho_{e}-ce$, $z_s=\rho_{s}-cs$, $z_a=\rho_{a}-ca$, $z_I=\rho_{I}-cI$, and sends $c, B_1,B_2, z_{r_1},z_{r_2},z_{\delta_1},z_{\delta_2}, z_{e},z_{s},z_{a}, z_{I}$ to the provider.
  \item The service provider computes $T'_1=B_1^c g_1^{\rho_{r_1}}g_2^{\rho_{r_2}}$, $T'_2=B_2^{-z_{e}} g_1^{z_{\delta_1}}g_2^{z_{\delta_2}}$, $T'_3=(\frac{\hat{e}(B_2,T)}{E_0})^c\hat{e}(B_2,g)^{-z_{e}}E_1^{z_{s}}E_2^{z_{a}}E_3^{z_{I}}E_4^{z_{r_1}}E_2^{z_{\delta_1}}$ and accepts if $c=\mathcal{H}(T_1,T_2,T_3,R)$; otherwise, rejects.
\end{enumerate}

\noindent  $\mathcal{PK}_3\{(A_i,e_i,s_i,a_i, I_i): \hat{e}(A_i,Tg^{e_i}){=}\hat{e}(g_0g_1^{s_i}g_2^{a_i}g_3^{I_i},g)\}$ is the same as $\mathcal{PK}_2$

$$\mathcal{SPK} \left \{ \begin{array}{c}  (B_i,f_i,t_i,t'_i, a_i,I_i,P_i,v_i):~~~~~~~~~~~~~~~~~~~ \\ ~~~~~~~~\hat{e}(B_i,Th^{f_i}){=}\hat{e}(h_0h_1^{t_i}h_2^{a_i}h_3^{I_i}h_4^{P_i},h) \land \\ ~~~~~~~~C'_i=h_1^{t'_i}h_2^{a_i}h_3^{I_i}h_4^{P_i}\land\\
~~~~~~~~P_i>Q_i \land\\
~~~~~~~~Y_i=H^{v_i}\land \\
~~~~~~~~Z_i=\hat{e}(g,\widehat{A}_i)\mathcal{G}^{X_iv_i} \end{array}\right \}(num).$$

\begin{enumerate}
  \item The mobile user randomly chooses $r_1,r_2,r_3,r_4 \in \mathbb{Z}_p$ to compute $B_1=y_1^{r_1}y_2^{r_2}$, $B_2=By_1^{r_2}$, $B_3=y_1^{r_3}y_2^{r_4}$, $B_4=\phi_{P-Q}y_1^{r_4}$. The mobile user picks random $\rho_{r_1},\rho_{r_2},\rho_{r_3},\rho_{r_4}, \rho_{f},\rho_{t}, \rho_{t'},\rho_{a},\rho_{I}, \rho_{P}, \rho_{v},\rho_{\omega_1},\rho_{\omega_2},\rho_{\omega_3},$ $\rho_{\omega_4}\in \mathbb{Z}_p$ and computes
      $T_1=h_1^{\rho_{t'}}h_2^{\rho_{a}}h_3^{\rho_{I}}h_4^{\rho_{P}}$, $T_2=y_1^{\rho_{r_1}}y_2^{\rho_{r_2}}$, $T_3=B_1^{-\rho_{f}}y_1^{\rho_{\omega_1}}y_2^{\rho_{\omega_2}}$, $T_4=K_0^{\rho_{r_2}}K_1^{\rho_{\omega_2}}F_1^{\rho_{t}}F_2^{\rho_{a}}F_3^{\rho_{I}}F_4^{\rho_{P}}\hat{e}(B_2,h)^{-\rho_{f}}$, $T_5=y_1^{\rho_{r_3}}y_2^{\rho_{r_4}}$,
      $T_6=B_3^{-\rho_{P}}y_1^{\rho_{\omega_3}}y_2^{\rho_{\omega_4}}$,
      $T_7=K_2^{\rho_{\omega_3}}K_3^{\rho_{\omega_4}}\hat{e}(B_4,y)^{-\rho_{P}}$,
      $T_8=H^{\rho_{v}}$,
      $T_9=\mathcal{G}^{X\rho_{v}}$.
  \item The user computes $c=\mathcal{H}(T_1,T_2,T_3,T_4,T_5,T_6,T_7,T_8,T_9,num)$.
  \item The user computes $z_{r_1}=\rho_{r_1}-cr_1$, $z_{r_2}=\rho_{r_2}-cr_2$, $z_{r_3}=\rho_{r_3}-cr_3$, $z_{r_4}=\rho_{r_4}-cr_4$,
      $z_{f}=\rho_{f}-cf$, $z_t=\rho_{t}-ct$, $z_{t'}=\rho_{t'}-ct'$, $z_a=\rho_{a}-ca$, $z_I=\rho_{I}-cI$, $z_P=\rho_{P}-cP$, $z_v=\rho_{v}-cv$, $z_{\omega_1}=\rho_{\omega_1}-cr_1\omega_1$, $z_{\omega_2}=\rho_{\omega_2}-cr_2\omega_2$, $z_{\omega_3}=\rho_{\omega_3}-c(P-Q)\omega_3$, $z_{\omega_4}=\rho_{\omega_4}-c(P-Q)\omega_4$,  and sends $c, B_1,B_2,B_3,B_4,z_{r_1},z_{r_2},$ $z_{r_3},z_{r_4}, z_{f},z_{t}, z_{t'},z_{a},z_{I}, z_{P}, z_{v},z_{\omega_1},z_{\omega_2},z_{\omega_3},z_{\omega_4}$ to the service provider.
  \item The service provider computes $T'_1={C'}^c h_1^{z_{t'}}h_2^{z_{a}}h_3^{z_{I}}h_4^{z_{P}}$, $T'_2=B_1^c y_1^{z_{r_1}}y_2^{z_{r_2}}$,
      $T'_3=B_1^{-z_{f}}y_1^{z_{\omega_1}}y_2^{z_{\omega_2}}$,
      $T'_4=(\frac{\hat{e}(B_2,T)}{F_0})^cK_0^{z_{r_2}}K_1^{z_{\omega_2}}F_1^{z_{t}}F_2^{z_{a}}F_3^{z_{I}}F_4^{z_{P}}\hat{e}(B_2,h)^{-z_{f}}$,
      $T'_5=B_3^cy_1^{z_{r_3}}y_2^{z_{r_4}}$,
      $T'_6=B_3^{-Qc}B_3^{-z_{P}}y_1^{z_{\omega_3}}y_2^{z_{\omega_4}}$,
      $T'_7=(\frac{\hat{e}(B_4,{\eta}y^{-Q})}{K})^cK_2^{z_{\omega_3}}K_3^{z_{\omega_4}}\hat{e}(B_4,y)^{-z_{P}}$,
      $T'_8=Y^cH^{z_{v}}$,
      $T'_9=(\frac{Z}{\hat{e}(g, \widehat{A})})^c \mathcal{G}^{X z_{v}}$,
  and accepts the proof if $c=\mathcal{H}(T'_1,$ $T'_2,T'_3,T'_4,T'_5,T'_6,T'_7,T'_8,T'_9,num)$; otherwise, rejects.
\end{enumerate}



\vfill\eject

\end{document}